\documentstyle[aps]{revtex}

\newcommand{\dba}{\not{\!{\!D}}}

\newcommand{\Abar}{\not{\!{\!A}}}
\newcommand{\pabar}{\not{\!{\partial}}}
\newcommand{\Od}{{\cal O}}

\newcommand{\tr}{\mbox{tr}}
\newcommand{\Tr}{\mbox{Tr}}

\newcommand{\Dbar}{\not{\!{\!D}}}

\newcommand{\omegabar}{\not{\!\omega}}

\newcommand{\intt}{\int_T d^2x }

\newcommand{\dmd}{\Dbar^\dagger(A;\mu) \Dbar(A;\mu)}
\newcommand{\ddm}{\Dbar(A;\mu) \Dbar^\dagger(A;\mu)}

\newcommand{\sgn}{\mbox{sgn}}
\newcommand{\diag}{\mbox{diag}}

\def\IN{\relax{\rm I\kern-.18em N}}
\def\IR{\relax{\rm I\kern-.18em R}}
\def\ID{\relax{\rm I\kern-.18em D}}

\font\cmss=cmss12 \font\cmsss=cmss12 at 7pt
\def\IZ{\relax\ifmmode\mathchoice
{\hbox{\cmss Z\kern-.4em Z}}{\hbox{\cmss Z\kern-.4em Z}}
{\lower.9pt\hbox{\cmsss Z\kern-.4em Z}}
{\lower1.2pt\hbox{\cmsss Z\kern-.4em Z}}\else{\cmss Z\kern-.4em Z}\fi}

\def\inbar{\,\vrule height1.5ex width.4pt depth0pt}
\def\IC{\relax\hbox{$\inbar\kern-.3em{\rm C}$}}

\newcommand{\NP}[1]{{\it Nucl.\ Phys.\ }{\bf #1}}
\newcommand{\ZP}[1]{{\em Z.\ Phys.\ }{\bf #1}}
\newcommand{\RMP}[1]{{\em Rev.\ of Mod.\ Phys.\ }{\bf #1}}
\newcommand{\PL}[1]{{\em Phys.\ Lett.\ }{\bf #1}}

\newcommand{\AN}[1]{{\em Ann. Phys. }{\bf #1}}
\newcommand{\CMP}[1]{{\em Comm.\ Math.\ Phys.\ }{\bf #1}}
\newcommand{\PRep}[1]{{\em Phys.\ Rep.\ }{\bf #1}}
\newcommand{\PR}[1]{{\em Phys.\ Rev.\ }{\bf #1}}
\newcommand{\PRL}[1]{{\em Phys.\ Rev.\ Lett.\ }{\bf #1}}

\newcommand{\IJmp}[1]{{\em Int.\ J.\ Mod.\ Phys.\ }{\bf #1}}

\begin{document}
\begin{flushright}
FT/UCM/2-97

IMPERIAL/TP/96-97/44

(to appear in {\it Phys.Rev.}D)
\end{flushright}
\vspace{.75cm}
{\large\bf\begin{center}
 The Schwinger and Thirring models at finite chemical 
potential and temperature
\end{center}} 
\vspace{.25cm}
%\maketitle

%\vspace{-1cm}
\begin{center}

{Ram\'on F.Alvarez-Estrada}\footnote{E-mail: 
ralvarez@eucmax.sim.ucm.es}\\{\it Departamento de F\'{\i}sica 
Te\'orica, Universidad Complutense,  28040, Madrid, Spain}\\ 
\hfill \\ 
 {Angel G\'omez Nicola}\footnote{E-mail: 
a.gomez@ic.ac.uk. 
On leave of absence from {\it Departamento de F\'{\i}sica 
Te\'orica, Universidad Complutense, 28040, Madrid, Spain}}
\\{\it Theoretical Physics Group, Imperial College, Prince 
Consort Road,  London SW7 2BZ, United Kingdom}
\vspace{.25cm}

(October, 29, 1997)
\end{center}
\vspace{.5cm}
\begin{abstract}
 The imaginary time generating functional $Z$ for the
 massless Schwinger model 
at nonzero chemical potential $\mu$ and temperature 
 $T$ is studied in a torus with spatial length $L$. 
The lack 
of hermiticity of the Dirac operator gives rise to a 
 non-trivial $\mu$ and $T$ dependent phase ${\cal J}$  
in the effective action. When
the Dirac operator has no zero modes (trivial sector), 
we evaluate ${\cal J}$, which is 
a topological 
contribution, and  we  find  exactly $Z$,   
the  thermodynamical partition function, the boson propagator and 
 the thermally averaged Polyakov loop. 
 The  $\mu$-dependent contribution of  the free partition 
function 
cancels exactly the nonperturbative one from ${\cal J}$, 
  for $L\rightarrow\infty$, yielding a zero  charge 
density for the system,
 which bosonizes at nonzero $\mu$. The boson 
 mass is $e/\sqrt{\pi}$, independent on $T$ and $\mu$, which is also 
 the inverse correlation length between two opposite charges.  
 Both the  boson propagator and the Polyakov loop  acquire
 a new $T$ and $\mu$ dependent 
 term at $L\rightarrow\infty$. The imaginary time generating functional 
 for the massless Thirring model at nonzero $T$ and $\mu$ 
is obtained exactly 
 in terms of the above solution of the 
 Schwinger model in the trivial sector. 
For this model, the $\mu$ dependences of the 
thermodynamical partition 
function, the total fermion number density and the fermion two-point 
 correlation  function are obtained. The phase ${\cal J}$ displayed 
  here leads to our new results and allows to complement non 
trivially previous
studies on those models. 
\end{abstract}

\vspace{2cm}

\newpage

\section{Introduction}

The Schwinger model is QED in 1+1 space-time dimensions 
\cite{sch62}. Although 
it is a toy model, it shares many interesting physical 
properties with more realistic theories 
such as QCD or the electroweak theory. 
 It is perhaps the simplest example in which gauge invariance 
does not necessarily 
imply a massless gauge boson, analogously to the Higgs 
phenomenon.  
 Other interesting properties of the model are dynamical 
mass generation, 
chiral symmetry breaking and confinement. 
The model with massless fermions was shown to be  exactly 
solvable in vacuum 
(that is, without thermal effects) long time ago 
\cite{sch62,losw71}. 
It is equivalent to a theory describing a free boson  with 
 mass $e/\sqrt{\pi}$ (bosonization), which is physically 
a fermion-antifermion 
bound state (confinement). 
The finite mass implies a finite correlation  length, 
which physically 
corresponds to charge screening, long range 
forces being  absent.  On the other hand, chiral symmetry 
is broken through 
the chiral anomaly \cite{john63} rather than spontaneously, 
 since  Coleman's theorem \cite{col73} prevents any continuous 
symmetry to be 
spontaneously broken in two dimensions. 
 When a mass parameter for the 
fermions is included, the model is no longer solvable but 
it is still possible 
 to analyse exactly  some of 
 the above properties, like fermion confinement \cite{coja75}.  

 In the last ten years, there has been a renewed interest in 
the study of 
the 
Schwinger model including statistical mechanics features. 
The  model was also 
solved and,
in particular, the two point correlation 
functions and the partition function were obtained at 
finite temperature $T$ and zero chemical potential
 $\mu$ in \cite{rual87} 
when  the Dirac operator has no zero modes. 
  Its thermodynamics can be expressed in terms of 
  those of a free boson of mass $e/\sqrt{\pi}$ and  
free fermions,
 i.e., bosonization also takes place at finite temperature.
 A more complete 
study of the 
 Schwinger model on a torus, which naturally incorporates 
 temperature effects, 
still at $\mu=0$, has been performed 
 in \cite{sawi92,sb9495}. More specifically,in  
\cite{sawi92} the model was 
treated with an 
arbitrary number of zero modes and the two-point fermion 
correlation function 
was calculated, whereas in \cite{sb9495} 
 higher  correlation functions were obtained. 
  In \cite{chsch96}, the correlation functions have also 
been studied for
  nonzero $\mu$.    In a recent work 
\cite{griso96}, the problem of charge screening 
at finite temperature (with $\mu=0$) in the Schwinger model 
is analysed, in 
connection with the 
spontaneous breaking of the discrete $Z$ symmetry, which 
corresponds to the 
freedom of choosing 
 gauge fields in the Euclidean time direction with any 
winding number around 
$S^1$. These  nontrivial  gauge transformations will play an 
essential role in our analysis with a 
nonzero chemical potential.   

On the other hand, we remind that the Thirring model, which 
describes massless 
fermions in 1+1 dimensions with a quartic 
 self-interaction, can also be explicitly solved in vacuum 
($T=\mu=0$) 
\cite{thi58,fried72}. We also recall that the generating 
 functional for the Thirring model at finite $T$ and $\mu=0$ 
has been obtained 
in terms of the fermionic one (with an external 
 electromagnetic source) for the Schwinger model \cite{rual87}. 
The Thirring 
model at nonvanishing $T$ and $\mu$  
 has been analysed in \cite{yoko87} for real time and in 
\cite{sawi96} in the 
torus.

 In this paper, we shall study, first, the Schwinger model in 
a medium at 
thermodynamical equilibrium, 
by introducing both the temperature and the fermion chemical 
potential 
$\mu$. 
 By considering a  nonzero $\mu$, we are able  
to study the system 
 when there is a finite net fermion charge density 
(in the free case, 
$\mu$ is 
just the Fermi energy). 
 Some of the  questions that naturally arise are: i) whether one 
can provide
simple solutions for the Schwinger model  
 with a 
non zero $\mu$, which extend  previous studies non trivially,
ii) whether bosonization takes place at finite fermion charge  density and,
 if 
so, which is the boson  mass,  
iii) which is   the net fermionic charge of the resulting system, that is, 
 whether  the fermions are still confined into neutral mesons  
 and iv) how does the chemical potential affect charge screening. We shall
 try to give 
  answers to all of these questions. The second aim of this work is to 
provide 
an exact solution for the Thirring
 model at $T\neq 0$ and $\mu\neq 0$ in the imaginary time formalism, in 
terms 
of the corresponding one for the Schwinger model, to compare with 
previous
findings by  other authors, using different methods, as a search for
consistency and to get some new results.

The plan of this paper is as follows. In section \ref{genfunden}, we shall 
deal with  the generating functional $Z$ of the 
 Schwinger model 
at nonzero $T$ and $\mu$, analysing  several important items: the 
fermionic generating functional $Z_f$ 
with an external electromagnetic field, the role of the zero modes, the
determinant of the Dirac operator, etc, by
following steps similar to those in \cite{sawi92}. The lack 
of hermiticity of the Dirac operator  and a non-trivial phase factor 
${\cal J}$ 
will be  genuine and 
crucial features of the $\mu\neq 0$ case. They both will make necessary 
 to extend the methods developed in \cite{sawi92}.  
   From section \ref{genfuntri} onwards, we shall restrict ourselves 
to the trivial sector, which is the only relevant one in
order to study the 
thermodynamics of the system. 
 We shall get  $Z$, $Z_f$ and ${\cal J}$ by using functional methods, 
  generalising what was done for 
 $\mu=0$ in \cite{rual87} and deriving the proper extension of  the 
point-splitting 
regularization when $\mu\neq 0$.   
Section \ref{phyres} is devoted to: a) 
several 
physical results for the Schwinger 
 model: the fermion charge density,  the thermodynamical partition 
function, 
  the boson propagator in the trivial sector,  
 the Polyakov loop (the order parameter of the confining symmetry) 
 and the screening length, b)
  the consistency of our methods. 
 The tasks of obtaining an explicit solution  and new results 
for the Thirring model at nonzero 
$T$ and $\mu$
 are undertaken in section  \ref{thirring}. Section \ref{conc}
 contains the 
conclusions and some discussions. Several 
  results pertaining to the zero modes in the 
Schwinger model at 
nonzero $T$ and $\mu$ are collected in an 
 Appendix.   

\section{The generating functional at finite temperature and density}
\label{genfunden}

Our starting point will be the generating functional for the
Schwinger model in the 
imaginary time formalism of Thermal Field Theory \cite{ber74,kap89}.
 We shall work in Euclidean 
two-dimensional space-(imaginary) time. In principle, we shall keep 
the length of the system $L$ as finite, 
 by imposing suitable boundary conditions in the spatial 
direction (see below). Thus, one
 properly defines 
the spectrum of the Dirac operator and  avoids infrared divergences 
 \cite{sawi92,sb9495,sawi96}. At the end of the calculations we shall 
take 
the $L\rightarrow\infty$ limit. The finite density  effects will be 
implemented by including a chemical potential $\mu$ associated to 
the conservation of the 
total electric charge (or the number of electrons minus that of 
positrons). Let
$A=(A^\mu)=(A^0, A^1)$ be the electromagnetic potential. Then, 
the generating functional reads
\begin{eqnarray}
Z[J,\xi,\overline\xi]&=&N(\beta)
 \int_{periodic}\!\!\!\!\!\!\!\!\!\!\!\!\!
 {\cal D} A \exp\left[ \intt \left(\Gamma[A]
+JA\right)\right]Z_f[A,\xi,\overline\xi]\nonumber\\
\Gamma[A]&=&-\frac{1}{2}E^2-\frac{1}{2\alpha}(\partial_\mu A^\mu)^2
\label{genfun}
\end{eqnarray}
where the fermionic generating functional is
\begin{equation}
 Z_f[A,\xi,\overline\xi]= \int_{antiperiodic} 
\!\!\!\!\!\!\!\!\!\!\!\!\!\!\!\!\!\!\!
{\cal D}\overline\psi {\cal D}\psi
 \exp \left[\intt \left(-\overline\psi \dba(A;\mu) 
\psi+\overline\xi\psi+\overline\psi\xi\right)\right] 
\label{fergenfun}
\end{equation}
and the Dirac operator is given by
\begin{equation}
\Dbar(A;\mu)=\pabar-ie\Abar-\mu\gamma^0
\end{equation}

In the above equations, $N (\beta)$ is a temperature dependent 
normalisation 
constant, $\beta=1/T$, $T$ being the temperature, $\int_T$ is the 
integral over the 
Euclidean two-dimensional torus $[0,\beta]\times [0,L]$ and $e$ is the
 electric charge, which has dimensions of energy. 
The fermionic and bosonic external sources are 
$\overline\xi$, $\xi$ and $J$ respectively. 
 The electric field is  
$E=F_{01}=\partial_0 A_1-\partial_1 A_0$ and $\alpha$ is the covariant 
gauge-fixing parameter. It is important to remark 
 here that the above covariant 
gauge fixing does not fix the  gauge completely on the torus. 
There is still 
some residual 
 gauge arbitrariness related to global gauge transformations, 
which we shall 
   deal with later.  
The Faddeev-Popov determinant has been absorbed 
in the  measure in (\ref{genfun}), as it  plays 
no dynamical role. 
Our conventions for the Euclidean Dirac matrices 
 ($\{\gamma_\mu,\gamma_\nu\}=
\delta_{\mu\nu}$) are:
$\gamma^0=\gamma_0$, $\gamma^1=\gamma_1$ and 
$\gamma^5=-i\gamma^0\gamma^1$ are the Pauli matrices.

 The electromagnetic field  
and the bosonic external 
source   are periodic in the 
Euclidean time with period $\beta$ whereas the fermionic 
fields and 
sources are anti-periodic. An alternative approach, which we 
shall not 
 follow here, would have been to take $\Dbar (A;0)$ in 
(\ref{fergenfun}), 
 with fermions satisfying the boundary condition 
$\psi (x^0+\beta,x^1)=-\exp (\beta\mu)\psi(x^0,x^1)$. 
Concerning the spatial boundary 
 conditions, they cannot be chosen as periodic, 
in general (after the above
 choice for 
the  temporal ones), as the 
Dirac operator may have zero modes on the torus 
 (to avoid duplications, we refer to \cite{sawi92} for a
justification).  Without loss of generality,  
we shall choose $A_\mu$ so that 
$A_\mu(x^0,x^1+L)-A_\mu (x^0,x^1)=\partial_\mu (-\Phi x^0/e\beta)$ 
and hence

\begin{eqnarray}
\psi (x^0,x^1+L)&=&\exp\left(-i\frac{\Phi}{\beta}x^0\right)\psi(x^0,x^1)
\nonumber\\
\overline\psi (x^0,x^1+L)&=&\overline\psi(x^0,x^1)
\exp\left(i\frac{\Phi}{\beta}x^0\right)
\label{ferspbc},
\end{eqnarray}
 with  $\Phi$ the total 
flux of the electric
field over the torus
\begin{equation}
\Phi=e\intt E(x)=2\pi (n_+-n_-)
\label{axanom}
\end{equation}
 where  $n_{\pm}$ are the number of   zero modes with positive and 
negative chirality. 
 The relation (\ref{axanom})  follows directly from the axial anomaly  
\cite{sawi92,fuji80}. We shall introduce $k=n_+ + n_-$, the total 
number of zero modes. The 
 gauge sector with $k=0$ will be referred to as the trivial sector.   
 For later use, we recall the following factorisation property of 
the Dirac operator
\begin{equation}
\Dbar (A;\mu)=\exp (x^0 \mu) \Dbar (A;0) \exp(-x^0 \mu)
\label{facprop}
\end{equation}

\subsection{General structure of $Z_f[A,\xi,\overline\xi]$ with 
zero modes}
\label{factzm}

 The contribution of the zero modes 
 to the generating functional  has to be analysed carefully, in 
order to properly define the functional determinant of the Dirac
 operator. 
For that purpose, we shall follow the same steps as in 
\cite{sawi92}. However, there is an important distinctive feature 
of the 
 $\mu\neq 0$ case, namely, that $i{\Dbar}[A;\mu]$ is non-hermitian.
 Hence, the 
set of  eigenfunctions of $i{\Dbar}$ is no longer 
an orthonormal basis in which the spinor fields could be expanded. 
To avoid 
this difficulty we shall expand the spinors in the basis 
of the hermitian operators ${\Dbar}^\dagger{\Dbar}$ 
and ${\Dbar}{\Dbar}^\dagger$. 
 This  will allow  to separate the zero mode contribution up 
to a phase factor. 
 We shall discuss below  this factor and its relevance to the 
calculation. First, let us consider the set of  eigenfunctions of 
the hermitian 
operators
\begin{equation}
 H(A;\mu)\phi_n=\left[\dmd\right]\phi_n=\mu_n \phi_n 
\quad ; \quad 
\overline H(A;\mu)\varphi_n=\left[\ddm\right]\varphi_n=\mu_n \varphi_n
\label{hhbar}
\end{equation}

The operators $H$ and $\overline H$ have the same eigenvalues 
 $\mu_n \geq 0$ (for $\mu_n> 0$,  
 ${\Dbar} \phi_n$ is an eigenstate of $\overline H$) 
and  the zero modes of $H$ ($\overline H$) are the same as 
those of $\Dbar$ ($\Dbar^\dagger$). In addition, since 
the anomaly (\ref{axanom}) is $\mu$-independent (for general results 
on the 
 independence of anomalies on thermal effects, see \cite{anomftd}),  
 $n_+-n_-$ is the same for both $\Dbar (A;\mu)$ and $\Dbar (A;0)$. 
As 
  we shall see in section \ref{instanton},  all zero modes have 
always the same  chirality. Therefore, the number $k$ of zero 
 modes is the same for $H$, $\overline H$ and $\Dbar (A;0)$.

At this point let us expand the spinor fields $\psi$ and 
$\overline\psi$ as
\begin{eqnarray}
\psi (x)&=&\sum_{p=1}^k \alpha_p \phi_p + 
\sum_{q=k+1}^{\infty} \beta_q \phi_q \nonumber \\
\overline\psi (x)&=&\sum_{p=1}^k \overline\alpha_p 
\varphi_p^\dagger + \sum_{q=k+1}^{\infty} \overline\beta_q 
\varphi_q^\dagger
\end{eqnarray}
with
\begin{equation}
\varphi_q=\frac{1}{\sqrt{\mu_q}} \Dbar \phi_q 
\qquad q=k+1,\dots ,\infty
\label{chbasdmd}
\end{equation}
 $\phi_p$ ($\varphi_p$) being the zero modes of $H$ 
($\overline H$).
 In this basis, we have $(\varphi_q,\varphi_{q'})=
(\phi_q,\phi_{q'})=\delta_{q q'}$, where 
$(\chi,\psi)=\intt \chi^\dagger \psi$ is the scalar 
product on the torus.  We get for the fermionic action
\begin{equation}
\intt \overline\psi \Dbar(A;\mu)\psi=
\sum_{q=k+1}^{\infty}\sqrt{\mu_q}\overline\beta_q \beta_q
\end{equation}

Then, the action is diagonal in this basis and, by doing 
the integration over the grassmanian variables $\alpha_p$, 
the 
contribution of the zero modes can be factorized. 
A crucial point should be 
noticed here. As the 
 spinors $\psi$ and $\overline\psi$ are expanded in 
different basis, the Jacobian of the change of basis 
from ${\cal D}\psi$ to $\prod_{p,q}d\alpha_p d\beta_q$
 is not the inverse of that from 
   ${\cal D}\overline\psi$ to 
$\prod_{p,q}d\overline\alpha_p d\overline\beta_q$. 
This fact was already noted by 
Fujikawa \cite{fuji84} in the context of anomalies 
with non-hermitian Dirac operators. Since both changes 
of variables are 
formally unitary, when  doing them simultaneously  
we are left with some phase factor $\exp[i{\cal J}(A;\mu)]$. 
Notice that ${\cal J}(A;0)=0$ since then 
$H=\overline H=-{\Dbar}^2 (A;0)$. Also, in principle,
 the phase factor is 
different for every $k$ sector, a feature to be reminded 
by means of a superscript $(k)$. 
Thus, performing the Gaussian Grassman integrals over 
$d\overline\alpha d\alpha d\overline\beta d\beta$ we get
\begin{eqnarray}
 Z_f[A,\xi,\overline\xi]&=&\exp 
[i{\cal J}^{(k)}(A;\mu)]\exp\left[-i\intt d^2 y\overline\xi (x) 
G(x,y,eA;\mu)\xi(y)\right]
\nonumber\\
&\times &\prod_{p=1}^k \intt  d^2 y \overline\xi (x) 
\phi_p (x)\varphi^\dagger_p (y)\xi (y) \sqrt{{\det}'  
H(A;\mu)}
\label{zfsepzero}
\end{eqnarray}
where $\det '$ is the functional determinant when the 
zero modes are omitted
(or factored out) and 
$G(x,y,eA)=\sum_{q=k+1}^\infty \frac{1}{\sqrt{\mu_q}}\phi_q (x) 
\varphi_q^\dagger (y)$ 
is the exact fermionic two-point function, satisfying the 
differential equation
\begin{equation}
\Dbar (A;\mu) G(x,y,eA;\mu)=\delta^{(2)} (x-y)-
\sum_{p=1}^{k}\varphi_p (x)\varphi_p^\dagger (y)
\label{exact2pdeq}
\end{equation}
%\begin{equation}
%P(x,y)=\sum_{p=1}^{k}\varphi_p\varphi_p^\dagger
%\end{equation}

The second term on the right hand side of the above 
equation is  the 
 projector onto the $\overline H$ zero mode subspace. 
 For simplicity, we have omitted a superscript $(k)$ 
in both $G(x,y,eA;\mu)$ and ${\det}' H(A;\mu)$. 
From (\ref{zfsepzero}) we see that the zero mode 
contribution can be 
factorized in this basis, in which we obtain 
 the contribution of  $\vert\det ' 
{\Dbar}\vert=({\det} ' H)^{1/2}$. However,
 we have still to
clarify which is the role of the phase factor $\exp(i{\cal J})$: 
this  will 
carried out in the next sections. Let us now remind  how
 to obtain different quantities of physical interest from 
(\ref{zfsepzero}). If we are interested in the thermodynamics 
of the 
 Schwinger model, the relevant quantity is the partition 
function $Z=Z(0,0,0)$, so that, from (\ref{zfsepzero}),  only the 
 trivial sector contributes:
\begin{equation}
Z(0,0,0)=N \int_{periodic,0}\!\!\!\!\!\!\!\!\!\!\!\!\!
 {\cal D} A \exp\left[ i{\cal J}^{(0)}(A;\mu)+\intt 
\Gamma[A]\right]\sqrt{\det H(A;\mu)}
\label{parfunfor}
\end{equation}  
We can obtain  thermodynamic observables such as the free 
energy and
 the particle charge  density, by differentiating $Z$ with 
respect 
to the temperature and the chemical potential respectively, 
thereby generalising for finite charge 
density the study carried out in 
\cite{rual87}. The subscript '0' in the functional integral 
above indicates 
that only the trivial sector contributes. Then,  in the 
trivial sector 
 ${\cal J}$ can be identified with the phase of the fermionic 
determinant. 
We can also calculate the average fermion charge density 
$\rho\equiv L^{-1}\int_0^L dx^1 
\langle\overline\psi\gamma^0\psi\rangle$  
 in terms of the two point Green function. A remark is in order 
 here: the equations of motion for the $A$ field imply 
 $\partial_1 E(x)=-ie\overline\psi\gamma^0\psi$, which is Gauss law. 
 If the latter is imposed as a quantum constraint  on  
  physical states \cite{jac80}, then it would imply 
 $\rho=0$  with the boundary conditions 
 chosen. As it is customary  \cite{mor78}, one may consider 
 that  an external compensating (say, ion) charge $\rho_{ex}$ is 
present to ensure 
 charge neutrality  
and hence Gauss law holds for the total charge 
 $\rho_{tot}=\rho+\rho_{ex}=0$. 
 Alternatively, one may  consider an open system 
 that exchanges particles with a reservoir ensuring charge neutrality. 
 With this in mind, we shall make no further reference to $\rho_{ex}$ 
and concentrate only in the fermion charge density $\rho$ for the 
 electron-photon system. Therefore,  (\ref{zfsepzero}) yields:
\begin{eqnarray}
\rho&=&\frac{1}{L}\int_0^L dx^1 
\langle\overline\psi\gamma^0\psi\rangle=
\frac{1}{\beta L}\frac{\partial}{\partial\mu}\log Z=
-\frac{1}{L}\int_0^L dx^1\frac{1}{Z}\frac{\delta}{\delta \xi}
\gamma^0\left.\frac{
\delta}{\delta\overline\xi}Z[0,\xi,\overline\xi]
\right\vert_{\xi=\overline\xi =0}\nonumber\\
&=&\frac{i}{Z}\frac{1}{L}\int_0^L dx^1\left\{
\int_{periodic,0}\!\!\!\!\!\!\!\!\!\!\!\!\!
 {\cal D} A \exp  \left[i{\cal J}^{(0)}(A;\mu)+\intt 
\Gamma[A]\right]\sqrt{\det H (A;\mu)}\right.\nonumber\\
&\times&\tr[\gamma^0 G(x,x)]  +i\int_{periodic,1}
\!\!\!\!\!\!\!\!\!\!\!\!\!
 {\cal D} A \exp \left[i{\cal J}^{(1)}(A;\mu)+\intt 
\Gamma[A]\right]
\left.\sqrt{{\det}'H(A;\mu)}\varphi_1^\dagger 
(x)\gamma^0 \phi_1 (x)\right\}
\end{eqnarray} 
and the  $k\geq 1$ contributions vanish. 
We shall show in section \ref{instanton} that the zero 
modes of $H$ and $\overline H$ have all the same chirality,   
given by $\sgn (\Phi)$. Hence, $\varphi_1$ and $\phi_1$
 are both eigenstates of $\gamma^5$ with the same eigenvalue and 
therefore the second piece in the above equation vanishes. 
Then, only the
 trivial sector contributes to the 
fermion number density.

 We also remark that the property  $\{\gamma_5,G(x,y)\}=0$, 
which is not difficult to prove with the above 
 definitions,  implies, 
 as in the case of finite $T$ but vanishing $\mu$ 
\cite{sawi92}, that the chiral 
condensate $\langle \overline\psi P_{\pm}\psi\rangle$, 
with $P_{\pm}=(1\pm\gamma^5)/2$, 
 does not depend on $G(x,y)$. However, from (\ref{zfsepzero})
 we see that it will contain  the phase 
 factor $\exp[i{\cal J}^{(1)}(A;\mu)]$ (see Appendix). 

\subsection{The imaginary part of the effective action}
\label{imagin}
%Note that we have formally 
% $Z_f^*(\overline\xi,\xi,{\Dbar})=
%Z_f(\xi,-\overline\xi,{\Dbar}^\dagger)$ and hence,  
%from (\ref{zfsepzero}) we have
%\begin{eqnarray}
%\vert  Z_f[A_\mu,\xi,\overline\xi]\vert&=&\frac{1}{2}\log 
%\dmd (A;\mu) + \frac{1}{2}\sum_{p=1}^k\left[ 
%\log \int d^2 x d^2 y \overline\xi (x) \phi_p (x)
%\varphi^\dagger (y)\xi (y)\right.\nonumber\\
%&+&\left.\log \int d^2 x d^2 y \overline\xi (x) 
%\phi_p^\dagger (x)\varphi (y)\xi (y)\right]-i
%\int d^2 x d^2 y\overline\xi (x) \re G(x,y)\xi(y)\nonumber\\
%\mbox{arg}Z_f[A_\mu,\xi,\overline\xi] &=& {\cal J}^{(k)}(A;\mu) 
%- \frac{i}{2}\sum_{p=1}^k\left[ 
%\log \int d^2 x d^2 y \overline\xi (x) \phi_p (x)
%\varphi^\dagger (y)\xi (y)\right.\nonumber\\
%&-&\left.\log \int d^2 x d^2 y \overline\xi (x) 
%\phi_p^\dagger (x)\varphi (y)\xi (y)\right]
%-i\int d^2 x d^2 y\overline\xi (x) \im G(x,y)\xi(y)
%\end{eqnarray}

 As  $i{\Dbar}\neq (i{\Dbar})^\dagger$, we have found  
the  extra factor ${\cal J}(A;\mu)$, which is the 
   source-independent piece of the phase of the 
generating functional. We shall analyse here its  physical 
interpretation,  at least in the trivial 
sector. The  general form of ${\cal J}(A;\mu)$ can be 
inferred 
  from the symmetry transformation properties of the 
phase of the different quantities  obtained from 
 $Z_f$ after switching off the external sources.  
  For instance, in the trivial sector, the object of 
interest is the effective action  $Z_f[A,0,0]$. 
 Now, recall that the 
$\mu$-dependent term in the Dirac action is odd under 
the operation of charge conjugation $C$, since it is 
the number of particles 
minus the number of antiparticles operator. The rest 
of the Dirac action is 
even under $C$, so that  $C$ acts   on $Z_f$ by 
replacing 
 ${\Dbar}\rightarrow -{\Dbar}^\dagger$ or, in 
 other words,  
$Z_f [A^C,0,0]=Z_f^* [A,0,0]$ and, therefore, 
the phase of the 
effective action is odd under $C$, while the modulus 
is even. This is analogous to the case of the QCD 
effective chiral lagrangian,  
when the symmetry under consideration is spatial 
parity ($P$), the 
phase of the effective action being, then,  
the Wess-Zumino-Witten term 
\cite{bij91}. 
 In our case, ${\cal J}^{(0)}(A;\mu)$ should contain 
only $C$-odd combinations of the gauge 
field. As $P$ is a symmetry of  
 the effective action, ${\cal J}$ should be $CP$ odd. In 
addition, it is not difficult to check that the $\mu$ 
term does not generate any 
anomaly  in  the gauge current, so that 
 imposing local gauge invariance (see comments below), 
 the only term which fulfils such symmetry requirements 
is of the form:
\begin{equation}
{\cal J}^{(0)}(A;\mu)= \tilde F(T,\mu,L)\intt A_0(x) 
\label{jterm}
\end{equation}
 where 
 $\tilde F(T,\mu,L)$ is a function, undetermined 
so far (to be found
explicitly later) such that  $\tilde F(T,\mu,L)=
-\tilde F(T,-\mu,L)$ 
since, by changing simultaneously $\mu\rightarrow 
-\mu$ 
and particle by antiparticle, the theory remains 
unchanged. There is another point that is worth 
noticing here. 
 Recall that in the torus  the gauge transformations $g(x^0,x^1):
S^1\times S^1\rightarrow U(1)$ are parametrised 
by $\IZ\times \IZ$, corresponding to 
the two winding numbers $(n,m)$ around the two circles.
 For any  $\Phi$,
  the most general gauge transformation  
    $A_\mu\rightarrow 
A_\mu + \partial_\mu \alpha$, which keeps $\partial_\mu A^\mu$ fixed 
 (so that $\partial_\mu \partial^\mu \alpha =0$) 
 and leaves unchanged the
 boundary conditions (in space and time) for both fermion and gauge 
fields 
is: $\alpha(x_0,x_1)= (2\pi n x^0)/\beta + (2\pi m x^1)/L$, up to 
 an additive constant. 
 Different choices of $(n,m)$  correspond to nontrivial, 
 homotopically disconnected, gauge transformations.  
  But then, we note that  the integral  in (\ref{jterm}) 
 is precisely equal to $n$  when $A$ is a pure gauge field. 
  Hence, (\ref{jterm}) 
changes by $n\tilde F$ when we perform a gauge 
transformation $g$ labelled 
 by $(n,m)$ and then it is not gauge invariant 
under  nontrivial,  gauge transformations. In this sense it is a 
topological term. 
 Therefore, we are imposing local gauge invariance 
but still allowing  a 
noninvariant topological term dependent 
 on the chemical potential. This assumption is 
partially motivated by previous
 works in which  $\mu$-dependent 
 topological effective actions were obtained  
\cite{rewi85,varios8595}. 
 In the next section, we shall get $\tilde F$ explicitly, through another
method, thereby 
 justifying our assumption.

\subsection{Instanton decomposition}
\label{instanton}
Following \cite{sawi92}, we shall 
 decompose the gauge field into an instanton part 
 $\tilde A_\mu$ and a local fluctuation $\phi$  as
\begin{equation}
A_\mu=\tilde A_\mu-\epsilon_{\mu\nu}\partial_{\nu}\phi
\label{deco1}
\end{equation} 
with $\phi (x)$ periodic in both space-time directions. 
 Note that  $\tilde A_\mu$  yields a constant electric 
field  
$\tilde E=\Phi/(eL\beta)$ and, hence, $E=\tilde 
E+\Delta\phi$. By following 
steps similar 
to those in \cite{sawi92}, we separate first 
the contribution of $\tilde A$ and
 $\phi$  in $\sqrt{\det' H}$. 
 Using that  ${\Dbar} (A;0)=
\exp\left(e\gamma^5\phi\right){\Dbar} (\tilde A;0) 
\exp\left(e\gamma^5\phi\right)$ 
 \cite{sawi92} and  (\ref{facprop}), it follows 
immediately that
\begin{equation}
{\Dbar} (A;\mu)=\exp(e\gamma^5\phi) \Dbar 
(\tilde A;\mu) \exp(e\gamma^5\phi)
\label{factalpha}
\end{equation}

Notice that the operator $H (A;\mu)$  in 
 (\ref{hhbar}) can be cast as
\begin{eqnarray}
H (A;\mu)&=&-\nabla_\nu \nabla_\nu (A,\mu)-
eE\gamma^5\nonumber\\  
\nabla_\nu&=&\partial_\nu-ie A_\nu 
+i\mu\gamma^5\delta_{\nu 1}
\label{h}
\end{eqnarray}
 and the operator $\overline H$ is obtained from 
$H$ by changing $\mu\rightarrow -\mu$. Hence, 
 $H(\tilde A;\mu)=-\nabla_\nu \nabla_\nu (\tilde A;\mu)
-\gamma^5\frac{\Phi}{L\beta}$, so that, as 
  $-\nabla_\nu \nabla_\nu$ in (\ref{h}) 
is a positive operator, all the zero modes 
of $H(\tilde A;\mu)$  have 
 the same  chirality, equal to the sign of $\Phi$ 
(recall that $[\gamma^5,H]=0$). On the other hand, from 
 (\ref{factalpha}), we get  a 
zero mode of
${\Dbar} (A;\mu)$ by  multiplying a zero mode of
 ${\Dbar} (\tilde A;\mu)$ by $\exp(-
e\gamma^5\phi)=\exp(-e\sgn (\Phi)\phi)$, which is in turn 
a zero mode of 
$H (A;\mu)$. We can apply exactly the same argument 
to $\overline H$.  Then, 
 the zero 
modes of $H$ and $\overline H$ in (\ref{hhbar}) 
have both the same chirality, 
which is equal to $\sgn (\Phi)$.   This 
was already used  in section \ref{factzm}, in order to omit the one 
zero mode contribution to the 
particle density.

It is possible to separate the contribution 
of ${\det} ' H(\tilde A;\mu)$ in 
${\det} ' H(A;\mu)$, for arbitrary $k$. We have sketched 
the derivation in the Appendix, the 
general formula, for any $k$, being given in (\ref{facinsta}). 
 From that expression, we read the usual induced  mass term for the  
boson field, with  mass $m=e/\sqrt{\pi}$, which is 
  independent 
on both the temperature and  the chemical potential, thereby 
generalising the 
result for $\mu=0$ previously derived in \cite{rual87,sawi92}. 
 Let us quote here the result  for the trivial sector 
 $k=0$:
\begin{eqnarray}
{\det} H(A;\mu)&=&{\det} H(\tilde A;\mu)\exp 
\frac{e^2}{\pi}\intt \phi (x) 
 \Delta \phi (x) 
\label{facinstatriv}
\end{eqnarray}

\subsection{The determinant of the instanton operator}

In order to complete the analysis in the previous section, 
one should still
study the spectrum of $H(\tilde A;\mu)$, which will be the 
purpose of 
the present section. 
First, by following \cite{sawi92}, we shall decompose the 
field $\tilde A$ as
follows: 
\begin{eqnarray}
\tilde A_0&=&-\frac{\Phi}{eL\beta}x_1+\frac{2\pi}{\beta} h_0+
\partial_0\lambda\nonumber\\
\tilde A_1&=&\frac{2\pi}{L}h_1+\partial_1\lambda
\label{deco2}
\end{eqnarray} 
which is the Hodge decomposition of the gauge field in the 
torus. The contributions proportional 
to $h_0$ and $h_1$ are the so called harmonic parts and are 
essential to correctly quantise the model 
 \cite{sawi92,sawi96}. Notice that under a nontrivial gauge 
transformation 
$(n,m)$ of the type commented in section \ref{imagin}, the $h$ fields 
 above are the only ones changing and they do so 
 as $h_0\rightarrow h_0+n$ and 
 $h_1\rightarrow h_1+m$, even for $\Phi=0$.  
The $\lambda$-dependent terms in the last two equations are pure gauge
contributions with $\lambda$  periodic in $x^0$ and $x_1$, 
which will not play any physical role. For instance, with the 
 covariant choice $\partial_\mu A^\mu=0$, $\lambda$ is just  a constant 
 and  that term does not appear in (\ref{deco2}). 
 Let us consider first the case   $\Phi=0$, which is the only relevant one 
 for the partition function.

Since $\gamma^5$ commutes with $\tilde H$, we choose the eigenfunctions of 
$ H(\tilde A;\mu)$ as states of definite chirality, that is: 
\begin{equation}
\Psi^+=\left(\begin{array}{c} \phi^+\\0 \end{array}\right)\qquad
\Psi^-=\left(\begin{array}{c} 0\\\phi^- \end{array}\right)
\end{equation}
with $\gamma^5\Psi^{\pm}=\pm\Psi^{\pm}$, since $\gamma^5=\diag (1,-1)$. 
Then, for $\Phi=0$ 
we have to solve 
$H^{\pm}(\tilde A;\mu) \phi^{\pm}=\lambda^{\pm}\phi^{\pm}$ 
with
\begin{equation}
 H^{\pm}(\tilde A;\mu)=
-(\partial_0-i\bar h_0)^2-(\partial_1-i\bar h_1\pm i\mu)^2
\end{equation}
 and  (anti-) periodic boundary conditions in the (time) space direction. 
In the above equation 
we have introduced $\bar h_0=2\pi e h_0/\beta$ and $\bar h_1=2\pi e h_1/L$. 
  The 
eigenfunctions are plane waves and the corresponding eigenvalues are:
\begin{equation}
\lambda^{\pm}_{nk}=
\left(\frac{2\pi}{\beta}\right)^2\left[n+\frac{1}{2}-eh_0\right]^2+
\left[\frac{2\pi}{L}(k-eh_1)\pm\mu\right]^2
\label{eigzeroflux}
\end{equation}
 with $n,k$ integers. Notice that 
the chemical potential breaks the chiral degeneracy which was
originally present in the $\mu=0$ case. Now, by using 
$\log\det H=\Tr\log H$ and (\ref{eigzeroflux}), we get:
\begin{eqnarray}
\log\det H(\tilde A;\mu)=
%\sum_{n,k=-\infty}^{\infty}\sum_\pm 
%\log \left[(\omega_n-\bar h_0)^2+(\omega_k-\bar h_1
%\pm\mu)^2\right]\nonumber\\
\frac{1}{2}\sum_{n,k=-\infty}^{\infty}\sum_{\pm\pm} 
\log \left[(\omega_n)^2+(\omega_k-\bar h_1\pm\mu\pm i\bar h_0)^2\right]
\end{eqnarray}
 with $\omega_n= (2n+1)\pi/\beta$ and $\omega_k=2\pi k/L$.
 Now, let us add and subtract $\log\beta^2\sum_{n,k}$ to the 
right-hand-side 
of
the above expression. This procedure will give rise to a $T$-dependent 
infinite constant, which, in turn,  will be 
absorbed, as customarily, in the normalisation constant $N(\beta)$ 
\cite{rual87,ber74,kap89}. We can perform the summation over $n$ in the 
above equation with the help of the two formulae \cite{kap89,gradryz80}: 
\begin{eqnarray}
\log\left[(2n+1)^2\pi^2+\beta^2(\omega_k\pm\alpha)^2
\right]&=&\int_1^{\beta^2(\omega\pm\alpha)^2}\frac{d\theta}{\theta^2+
(2n+1)^2\pi^2}\nonumber\\
&+&\log\left[1+(2n+1)^2\pi^2\right]\nonumber\\
\sum_{n=-\infty}^{\infty}\frac{1}{(2n+1)^2+\theta^2}&=&\frac{1}{\theta}
\left(\frac{1}{2}-\frac{1}{e^\theta+1}\right)
\end{eqnarray}
In so doing, we obtain, finally: 
\begin{eqnarray}
\log\det H(\tilde A;\mu)=\sum_{k=-\infty}^{\infty}\left\{2\beta
\left(\omega_k-\bar h_1\right)
+\sum_{\pm\pm} \log\left[1+\exp -\beta\left(\omega_k-\bar 
h_1\pm\mu\pm\bar h_0\right)\right]
\right\}
\label{parfunh}
\end{eqnarray}
up to an irrelevant $T$ and $\mu$ independent constant. 
 In order to obtain the full partition function, we have to multiply 
the above 
expression (which for   $e=0$ reproduces  the partition function for 
free fermions at finite density)  by $\exp(i{\cal J}^{(0)})$  
in (\ref{parfunfor})
 and by the $\phi$-dependent one in (\ref{facinstatriv}) and, then, 
integrate  
 over the gauge fields $(\phi,h_0,h_1)$. As a consequence of the 
decomposition of the gauge field chosen here, the phase factor 
 only depends on $h_0$, so that: 
\begin{equation}
\int_T d^2x A_0(x)=-\frac{\Phi L}{2e}+2\pi h_0 L
\label{phaseh0}
\end{equation}
since $\phi$ is periodic in the space direction. Then, the $\phi$ 
contribution to the partition function in (\ref{facinstatriv}) gives 
the partition function of a free massive boson \cite{rual87}. All the 
dependence on the chemical potential is included in the 
$(h_0,h_1)$ part, as given in (\ref{parfunh}) and  (\ref{phaseh0}).
 However, we have still  to determine  the value of 
$\tilde F(T,\mu,L)$ in (\ref{jterm}). 
Before undertaking that task, and for 
completeness, let us analyse the spectrum of 
 $H(\tilde A;\mu)$ when  $\Phi\neq 0$. In this case, we have
\begin{equation}
H^\pm (\tilde A;\mu)=-(\partial_0-i\bar h_0+i\frac{\Phi}{L\beta}x_1)^2-
(\partial_1-i\bar h_1\pm i\mu)^2
\mp e\tilde E
\label{hpmphinz}
\end{equation}
 together with the boundary conditions in the spatial direction 
given in (\ref{ferspbc}). This eigenvalue problem is solved in 
the Appendix. 
 From the result found there, we remark here 
that the $\mu$ dependence, when $\Phi\neq 0$, appears only in the 
states, while the determinant in (\ref{detpatilde})  depends on the 
temperature  
$T$ but not on $\mu$. As the norm of the zero modes in 
(\ref{normzeromod}) is 
also $\mu$-independent,  
then, if we go to (\ref{facinsta}) 
and (\ref{zfsepzero}), we realize that the dependence of the 
 fermionic generating functional on $\mu$, when $\Phi\neq 0$, 
is encoded in 
the 
determinants of the matrices in (\ref{facinsta}), that follow 
 immediately from the spectrum found for 
 $\alpha=0$. Besides, there are 
$\mu$ dependences in  both $G(x,y,eA;\mu)$ and ${\cal J}^{(k)}(A;\mu)$. 

\section{The generating functional in the trivial sector}
\label{genfuntri}
In this section, we shall use functional methods in order to 
calculate the 
generating functional in the trivial sector. 
 By employing these methods, we shall also obtain  the phase factor 
${\cal J}^{(0)}(A;\mu)$. 
This   will allow to get the full fermion charge  density
 and the partition function, as well as to establish the consistency 
with the 
results of the previous section. 
 We remind that 
 the covariant gauge-fixing is not complete on the torus. We still have  
 the freedom of performing a nontrivial gauge transformation 
 $(h_0,h_1)\rightarrow (h_0+n,h_1+m)$, with the $h$ fields in 
  (\ref{deco2}) and $n,m$ integers, 
corresponding to loops that 
wind $n$ times around the temporal direction and $m$ times around the 
spatial 
one. It is clear that  $n$ and $m$ are not fixed 
 by $\partial^\mu A_\mu =0$.  Throughout this section, though, 
we shall work  with the covariant gauge-fixing, ignoring this residual 
 gauge arbitrariness. The latter has also been treated in a situation 
related
to the one analysed here, but not quite identical with it: specifically, 
for
(real time) QED on a spatial circle, at zero temperature and chemical 
potential
\cite{He88}.  
 We have to bear in mind that 
 in (\ref{genfun}), we are integrating the gauge field over all possible 
 values of the fields $h_\mu$, that is, 
$h_\mu\in\IR$. Fixing the gauge for those fields would consist in 
 restricting them to  a range  $[0,1]\times\IZ$  \cite{sawi92}, since 
 they change by an integer under a global  gauge transformation. Then, 
 if the effective action is globally gauge invariant, and unaffected 
by the
residual gauge arbitrariness, the difference 
 between integrating over all $h$ or restricting  them 
 to a $[0,1]$ interval, 
 is an infinite constant independent on $T$ and $\mu$. So that 
 arbitrariness cannot affect physical observables 
 such as the free energy or the particle density. 
For $e=0$, the action depends on derivatives of the electromagnetic 
field, and
the covariant gauge fixing, even if not complete, suffices to get a
well-defined 
propagator, unaffected by that arbitrariness. In general, when $e$ is
non-vanishing, $Z[J,\xi,\overline \xi]$ is also unaffected by the 
arbitrariness,
after having integrated over all fields. However, for a given $A_\mu$, 
both
$\det H(\tilde A;\mu)$ and the fermionic generating functional $Z_f$ 
may be
subject to it, even if $\mu=0$. In particular, when $\mu\neq 0$, we have 
seen in section 
 \ref{imagin}  that $Z_f$ 
 is not globally gauge invariant, due to the induced topological term 
 in the phase ${\cal J}$, which  changes when $h_0\rightarrow h_0 + n$. 
 Thus, if we restrict $h_0$ to a $[0,1]$ interval, 
the result for the observables  would depend on our choice and then 
 it is consistent to let $h_0\in\IR$. We shall come again to this point 
 at the end of section \ref{charparfun}, where we shall  perform
   the integration 
 over the $h$ fields explicitly, using the results derived in 
 section \ref{genfunden}. It is not difficult to check 
 that all the formal functional manipulations that we shall carry 
 out in this section, except those related to $L[A]$, 
are also unaffected by 
the residual gauge arbitrariness.  

 Thus, as a first step, let us 
rewrite the generating functional in (\ref{genfun}) for the 
trivial sector,  
with the aid of standard functional techniques,  as: 
\begin{eqnarray}
Z[J,\xi,\overline\xi]&=&Z_{EM}Z_F\exp\left[-ie\intt 
\frac{\delta}{\delta\xi(x)}\gamma^\nu \frac{\delta}{\delta J^\nu(x)} 
\frac{\delta}{\delta\overline\xi(x)}\right]\nonumber\\
&\times&\exp\int_T d^2x d^2y\left[\frac{1}{2}J_\mu(x)
D^{\mu\nu}(x-y)J_\nu 
(y)-i\overline\xi (x)S(x,y;\mu)\xi (y)\right]
\label{funmetfirst}
\end{eqnarray}
where $Z_{EM}$ and $Z_F$ ($Z_F (T,\mu,L)=Z_F$ for short) 
are the free boson and fermion partition functions and 
$D^{\mu\nu}(x-y)$ and $S(x,y;\mu)$ are the free gauge boson and fermion 
propagators respectively:
\begin{eqnarray}
D_{\mu\nu}(x-y)&=&\frac{1}{\beta L}\sum_{n,k=-\infty}^{\infty}
e^{i\omega\cdot 
(x-y)}\frac{1}{\omega^2}\left[\delta_{\mu\nu} 
+(\alpha -1)\frac{\omega_\mu\omega_\nu}{\omega^2}\right]\nonumber\\
S(x,y;\mu)&=&-\frac{i}{\beta L}\sum_{n,k=-\infty}^{\infty}
e^{i\omega\cdot (x-y)}
\frac{1}{\gamma^0 (\omega_n+i\mu)+\gamma^1 \omega_k}
\label{freeprop}
\end{eqnarray}
 where $\omega=(\omega_n,\omega_k)$, with $\omega_k=2\pi k/L$, 
$\omega_n=2\pi n/\beta$ in the bosonic propagator and 
$\omega_n=(2n+1)\pi /\beta$ in the fermionic one. 
Now, we shall make use of 
some  known functional differentiation formulae \cite{fried72}, and, in
particular of: 
\begin{eqnarray}
\exp\left[-i\int d^2 x d^2 y \frac{\delta}{\delta\xi(x)}
A(x,y)\frac{\delta}
{\delta\overline\xi(y)}\right]\times
\exp i\int d^2 x d^2 y \overline\xi (x)B(x,y)\xi (y)\nonumber\\
=\exp\left[i\int d^2 x d^2 y\overline\xi (x) 
\overline B (x,y)\xi (y)+L\right]
\label{funmetthird}
\end{eqnarray}
Here, $A(x,y)$ and $B(x,y)$ are arbitrary functions, 
to be regarded as the 
kernels of the operators $A$ and $B$, 
respectively, $\overline B=B(1+AB)^{-1}$
 and $L=-\Tr\log [1+AB]^{-1}$, $\Tr$ indicating 
 the trace over functional and Dirac spaces.  Thus, one finds:

%\begin{eqnarray}
%Z[J,\xi,\overline\xi]=Z_{EM}Z_F\exp\left[\frac{1}{2}\int_T d^2 x d^2 y 
%J_\mu(x)D^{\mu\nu}(x-y)J_\nu (y)\right]\\
%\displaystyle
%\times\exp\left[ -\frac{1}{2}\int_T d^2 x d^2 y\frac{\delta}{\delta A_\mu 
%(x)}D^{\mu\nu}(x-y)\frac{\delta}{\delta A_\mu (y)}\right]
%\nonumber\\
%\times \exp\left[ e\intt \frac{\delta}{\delta\xi(x)}\gamma^\nu A_\nu 
%\frac{\delta}{\delta\overline\xi(x)}\right]
%\exp\left[-i\int_T d^2 x d^2 y \overline\xi (x)S(x,y;\mu)\xi (y)\right]
%\end{eqnarray}

\begin{eqnarray}
Z[J,\xi,\overline\xi]&=&Z_{EM}Z_F\exp\left[\frac{1}{2}\int_T d^2 x d^2 y 
J_\mu(x)D^{\mu\nu}(x-y)J_\nu (y)\right]\nonumber\\
&\times &\exp\left[ -\frac{1}{2}\int_T d^2 x d^2 y\frac{\delta}{\delta 
A_\mu (x)}D^{\mu\nu}(x-y)\frac{\delta}{\delta A_\mu (y)}\right]
\nonumber\\
&\times&\exp\left\{-i\int_T d^2 x d^2 y\overline\xi (x) G(x,y,ieA;\mu)\xi 
(y)+L[A]\right\}
\label{genfunfried}
\end{eqnarray}
with $A_\mu (x)\equiv -i\int_T d^2 y D_{\mu\nu}(x-y)J_\nu (y)$, 
after having
performed the functional differentiations, which appears to leave 
no trace of
the residual gauge arbitrariness in $Z[J,\xi,\overline\xi]$. 
 The so-called closed fermion loop functional $L[A]$ can be written  
formally 
as: 
\begin{equation}
L[A]=\tr_D\int_0^e de' \intt \Abar (x) G(x,x,ie'A;\mu)
\label{Lformal}
\end{equation}
 where $\tr_D$ denotes the Dirac trace. We remind that $G(x,y,eA;\mu)$  
is the two-point function, which, in the 
trivial sector, fulfils Eq. (\ref{exact2pdeq}) with $k=0$, that is, with 
its second term on the right hand side  omitted.

In order to get a well defined expression for the generating functional,
 we 
need, first, to regularise $L[A]$ in (\ref{Lformal}). 
For that purpose, we shall appeal to the  point-splitting regularization 
\cite{fried72}. Therefore, we should deal with the following limit: 
$\lim_{x\rightarrow y} G(x,y,eA;\mu)$. In Minkowski space-time, the limit 
should be taken by keeping the points $x$ and $y$ relatively 
space-like, in order to maintain causality \cite{fried72}. As we are 
working in
 Euclidean space-time, we shall not impose, in principle,  
such a restriction. We shall comment below on the different ways 
of taking the 
limit. Before that, and generalising \cite{fried72}, we 
shall derive the point-splitting regularization prescription 
in our present 
case with nonzero $\mu$. We start with the formal 
definition of the gauge current in the presence of an external 
background field
 $A_\mu$:
\begin{equation}
\langle j_\mu (x)\rangle_f [eA]=\langle\overline\psi (x)\gamma_\mu 
\psi(x)\rangle_f [eA] 
=i\lim_{x\rightarrow y} \tr_D \gamma_\mu G(x,y,eA;\mu)
\label{jformal}
\end{equation}
 where  $\langle O\rangle_f=\int {\cal D}\overline\psi 
{\cal D}\psi O 
\exp [-\int\overline\psi{\Dbar}\psi]$. 
 We shall obtain the regularised version of the right-hand-side of 
the above 
equation as follows: we shall demand that such a regularised gauge 
 current  be conserved and gauge invariant.
Notice 
that, under a gauge transformation 
$A_\mu\rightarrow A_\mu-\partial_\mu\Lambda$, $G(x,y,eA;\mu)$ changes as:
\begin{equation}
G(x,y,eA;\mu)\rightarrow G(x,y,eA;\mu)\exp ie[\Lambda(x)-\Lambda(y)]
\end{equation}

Based upon this, it is easy to show that the product: 
\begin{equation}
G(x,y,eA;\mu)\exp [-ie\int_x^y d\xi^\sigma A_\sigma (\xi)]
\label{almostreg} 
\end{equation}
is gauge invariant. However, if we replace (\ref{almostreg}) into 
(\ref{jformal}), calculate the divergence of the current $j_\mu$ 
so defined, 
and use  ${\Dbar} (A;\mu) G=\delta^{(2)} (x-y)$, we find that such a 
divergence does
 not vanish for $\mu\neq 0$. To ensure that the current 
is divergenceless, we have to add an extra $\mu$-dependent term, which 
leads to 
the regularised  gauge current:
\begin{equation}
\langle j_\mu (x)\rangle_f^{reg} [eA]=i\lim_{x\rightarrow y} \tr_D 
\gamma_\mu G(x,y,eA;\mu)\exp [-ie\int_x^y d\xi^\sigma A_\sigma (\xi)]
 \exp [-\mu (x^0-y^0)]
\label{jreg}
\end{equation}
which is, indeed, gauge invariant and divergenceless. 
Note that Euclidean covariance is broken since the system 
 is in a thermal bath. We are now ready 
to define the regularised fermion closed loop as:
\begin{eqnarray}
L^{reg}[A]&=&-i\tr_D\int_0^e de' \intt A_\mu (x)\langle j_\mu (x)
\rangle_f^{reg} [ie'A]\nonumber\\
&=&\tr_D\int_0^e de' \intt A_\mu (x) \lim_{x\rightarrow y} \tr_D 
\gamma_\mu G(x,y,ie'A;\mu)\nonumber\\
&\times &\exp [e'\int_x^y d\xi^\sigma A_\sigma (\xi)] 
\exp [-\mu (x^0-y^0)]
\label{Lreg}
\end{eqnarray} 

The limit 
$x\rightarrow y$ has to be taken in a symmetric way, regarding $(x,y)$ 
\cite{fried72}. 
In order to calculate the fermion closed loop  in  (\ref{Lreg}), 
 we shall consider an 
   ansatz for the exact Green function similar to that in \cite{rual87}:
\begin{equation}
G(x,y,eA;\mu)=\exp[-ie[\chi (x)-\chi(y)]S(x,y;\mu)
\label{gansatz}
\end{equation}

It is not difficult to check that, with the above ansatz, $G(x,y)$ is a 
solution of $\Dbar G(x,y)=\delta^{(2)}(x-y)$ , provided 
that $\chi (x)$ be 
a solution of ${\pabar}_x \chi (x)=-{\Abar} (x)$. 
 In turn, the solution is:
\begin{eqnarray}
\chi (x)&=&-\int_T d^2 y \Delta (x-y) \pabar_y \Abar (y)\nonumber\\
\Delta (x-y)&=&-\frac{1}{\beta L}\sum_{n,k=-\infty}^{\infty} 
e^{i\omega\cdot (x-y) }\frac{1}{\omega_n^2+\omega_k^2}
\label{solchi}
\end{eqnarray}
where  $\omega=(\omega_n,\omega_k)$ with $\omega_n=2\pi n/\beta$, 
$\omega_k=2\pi k/L$. So $G(x,y,eA;\mu)$ appears to be unaffected by the 
residual gauge arbitrariness.

Hence, from the above equations, we have to find the behaviour of 
$S(x,y;\mu)$ 
when $x\rightarrow y$. For that purpose, we apply 
the standard formulae:
\begin{eqnarray}
\frac{1}{\beta}\sum_{n=-\infty}^{\infty} 
f(\omega_n+i\mu)&=&\frac{1}{2\pi}\int_{-\infty}^{\infty} d\omega 
f(\omega)+
\oint_C d\omega f(\omega)\nonumber\\
&-&\frac{1}{2\pi}\sum_\pm\int_{-\infty\mp i\epsilon +i\mu}^{\infty\mp 
i\epsilon+i\mu}d\omega f(\omega) 
\frac{1}{e^{\pm i\beta\omega}+1}\nonumber\\
\frac{1}{\beta}\sum_{k=-\infty}^{\infty} f(\omega_k)&=&\frac{1}{2\pi}
\int_{-\infty}^{\infty} d\omega f(\omega)\nonumber\\
&+&
\frac{1}{2\pi}\sum_\pm\int_{-\infty\mp i\epsilon}^{\infty\mp 
i\epsilon}d\omega f(\omega) 
\frac{1}{e^{\pm i L\omega}-1}
\label{ftdformulae}
\end{eqnarray}
where $\epsilon\rightarrow 0^+$, $\omega_n=(2n+1)\pi i/\beta$, 
$\omega_k=2\pi k/L$ and  $C$ is the rectangular  contour in 
the complex $\omega$ plane running 
through the points $(+\infty,-\infty,-\infty+i\mu, +\infty+i\mu)$. 
We apply (\ref{ftdformulae}) to  $S(x,y;\mu)$ in 
(\ref{freeprop}) and retain only the dominant contributions in 
$x\rightarrow y$. The complex plane integrals along the lines 
$(-\infty\mp i\epsilon,+\infty\mp i\epsilon)$ and   
$(-\infty\mp i\epsilon+i
\mu,+\infty\mp i\epsilon+i\mu)$ are performed by forming
 a closed contour, by means of an infinite arc above (below), 
corresponding to 
the +(-) sign in (\ref{ftdformulae}), and applying the Residue Theorem.
  We get:
\begin{equation}
\lim_{x\rightarrow y} S(x,y;\mu)=e^{(x^0-y^0)\mu}\left[ \frac{1}{2\pi}
\frac{(x-y)_\mu\gamma^\mu}{(x-y)^2}+i\gamma^0 
F(T,\mu,L)+\Od (x-y)^2\right]
\label{limfreeprop}
\end{equation} 

where
\begin{equation}
F(T,\mu,L)=\frac{1}{2L}\sum_{k=-\infty}^{+\infty}
\left[\frac{1}{e^{\beta (\omega_k+\mu)}+1}-
\frac{1}{e^{\beta (\omega_k-\mu)}+1}\right]
\label{FLfin}
\end{equation}
with $\omega_k=2\pi k/L$. We recall that (\ref{limfreeprop})reproduces 
the $T=\mu=0$ 
result given in \cite{fried72} and the  $\mu=0$, $T\neq 0$ one 
($F(\mu=0)=0$) in  \cite{rual87}. 

 Next, we shall  replace both (\ref{solchi}) and the limit 
(\ref{limfreeprop}) into (\ref{gansatz}) 
and  (\ref{Lreg}).  We have taken  the  
$x\rightarrow y$ limit in two different ways and established that the same 
result is arrived at.  We have taken, firstly, 
$x^0-y^0\rightarrow 0$,  $x^1-y^1\rightarrow 0$ 
with $(x^1-y^1)/(x^0-y^0)=1$ 
and, secondly, the Minkowski causal choice 
(see \cite{fried72}) $x^0=y^0$ and $(x^1-y^1)\rightarrow 0$. Anyway, what
 it is important to note here  is that the exponential $\mu$ 
dependence in (\ref{limfreeprop}) is exactly cancelled with that in the 
regulator in (\ref{Lreg}). Then, no matter how we take the 
$x\rightarrow y$ limit, we always get a term 
$\tr\int {\Abar} (x)\gamma^0 F$ 
in 
$L^{reg}$. The possible divergence 
 in $L^{reg}$ arising from the first  piece in (\ref{limfreeprop}) 
is absent 
since we have taken  
the limit symmetrically. Thus, finally we arrive at
\begin{equation}
L^{reg}[A]=\frac{1}{2}\int_T d^2x d^2 y 
A_\alpha (x)\Pi^{\alpha\beta} (x-y) 
A_\beta (y)+2e F(T,\mu,L)\intt  A_0 (x)
\label{piexp}
\end{equation}
with
\begin{equation}
\Pi_{\alpha\beta} (x-y)=\frac{1}{\beta L}\frac{e^2}{\pi}
\sum_{n,k=-\infty}^{+\infty} e^{i\omega\cdot (x-y)}\left[
\delta_{\alpha\beta}-\frac{\omega_\alpha \omega_\beta}{\omega^2}\right]
\label{piprop}
\end{equation}
where $\omega_n=2\pi n/\beta$ and $\omega_k=2\pi k/L$. 
So, $L^{reg}[A]$ is, in
principle, 
affected by the residual gauge arbitrariness. The functional 
differentiations with respect to $A(x)$ in 
(\ref{genfunfried}) can be performed by employing the following formula, 
which 
is valid for any linear operators $P$ and $Q$ \cite{fried72}
\begin{eqnarray}
\exp\left[-\frac{i}{2}\int \frac{\delta}{\delta A }P 
\frac{\delta}{\delta A}\right]\exp\left[\frac{i}{2}
\int AQA+i\int f\cdot A\right]
=\exp\left[\frac{i}{2}\int A\overline P A\right.&&\nonumber\\
\left. +i\int A(1-QP)^{-1}\cdot f+\frac{1}{2}\tr\log (1-QP)^{-1}+
\frac{i}{2}\int f Q(1-QP)^{-1}f\right]&&
\label{recetafried3}
\end{eqnarray} 
where $\overline P=P(1-QP)^{-1}$ and we have omitted, for simplicity, 
all the space-time dependences.
We shall concentrate on the bosonic generating functional $Z[J,0,0]$, 
as we have already analysed the dependence on the fermionic 
sources in the previous section, up to the phase factor (to be 
calculated below). Upon applying   
 (\ref{recetafried3}) in (\ref{genfunfried}),  $L[A]$ being given in 
(\ref{piexp}), we get: 
\begin{eqnarray}
Z[J,0,0]&=&Z_{EM}Z_F\exp\left[\frac{1}{2}\int_T d^2 x d^2 y (J-iG)_\mu (x)
\ID_{\mu\nu}^{(0)} (x-y)(J-iG)_\nu (y)\right]\nonumber\\
&\times &\exp\left[ \frac{1}{2}\Tr\log (1+\Pi D)^{-1}\right]
\label{genfunJ}
\end{eqnarray}
where $G_0=2eF(T,\mu,L)$, $G_1=0$. In turn,  $\ID_{\mu\nu}^{(0)}$ is the 
exact boson propagator at $\mu=0$, which can be expressed, formally, as:
$\ID^{(0)}=D(1+\Pi D)^{-1}$, $D$ and $\Pi$ being given in (\ref{freeprop}) 
and 
(\ref{piprop}), respectively. Its explicit representation 
is: 
\begin{equation}
\ID_{\mu\nu}^{(0)} (x-y)=\frac{1}{\beta L}
\sum_{n,k=-\infty}^{\infty}e^{i\omega\cdot (x-y)} 
\left[\frac{1}{\omega^2+m^2}\left(\delta_{\mu\nu}-
\frac{\omega_\mu\omega_\nu}{\omega^2}\right)+
\alpha\frac{\omega_\mu\omega_\nu}{(\omega^2)^2}\right]
\label{ddprop}
\end{equation}
where $m^2=e^2/\pi$ is the induced boson mass. 

In the following section, we shall obtain, from (\ref{piexp}) and  
(\ref{genfunJ}),    
the complete form of the fermionic  generating functional 
in the trivial sector, including  the phase factor, the exact boson 
propagator at $\mu\neq 0$ and 
  the partition function  and check  the  results with  the method 
used in the previous section.  

\section{Physical results}
\label{phyres}
\subsection{The  charge density and the partition function} 
\label{charparfun} 
 By
 setting $J=0$ in the expression (\ref{genfunJ}) we get 
the partition function
\begin{equation}
Z(T,\mu)=\frac{Z_F(T,\mu)}{Z_F(T,\mu=0)}Z(T,\mu=0)\exp 
-\frac{1}{2}[2eF(T,\mu,L)]^2\int_T d^2 xd^2 y \ID_{00}^{(0)} (x-y)
\label{parfundd}
\end{equation}

However, the above integrals of $\ID_{00}$ are ambiguous, 
in the sense that 
 changes in the order in which such integrals are done yield 
different results. For instance, let us perform, firstly, the spatial
integrals:
 that would force us to set $k=0$ in 
(\ref{ddprop}), which would give rise, after doing the temporal integral, 
to an infinite and gauge dependent result. 
 Conversely, by changing the order of the integrals, 
let us perform, first, 
the one over the (imaginary) time. In so doing,  
 we arrive at a finite and gauge-independent answer. The latter 
prescription seems to be a natural and physically reasonable choice. In 
addition, as we shall see, it is consistent 
 with the results obtained in the previous section. If we 
adopt this prescription  in (\ref{parfundd}), we get:
\begin{equation}
Z(T,\mu)=\frac{Z_F(T,\mu)}{Z_F(T,\mu=0)}Z(T,\mu=0)\exp [-2\beta L\pi 
F^2(T,\mu,L)]
\label{parfunfin}
\end{equation}

This is a genuine nonperturbative 
result, since the argument of the exponential in the term that corrects 
the free fermionic partition function is independent of 
the electric charge. As we shall see, such a nonperturbative behaviour 
 comes 
directly from the topological structure 
discussed in section \ref{genfunden}, namely, from  
 the phase ${\cal J}[A;\mu]$ of the fermionic generating functional. 
At this point, 
we recall the result obtained in 
\cite{rual87} for $Z(T,\mu=0)$. The latter partition function 
was proven to 
factorise into the product of that 
for a free fermionic field times that for 
 a free massive boson field with mass $m$, divided by the one for a free 
massless field (finite temperature bosonization).  In our present case, 
with $\mu\neq 0$, we may wonder if such a 
factorisation actually takes place as well and, if so, whether the whole 
system may have 
a net fermionic charge or not. Let us consider, first, the free 
fermionic partition function $Z_F(T,\mu)$: 
\begin{equation}
\log Z_F(T,\mu,L)=
\sum_{k=-\infty}^{+\infty}
\left\{\log\left[ 1+e^{-\beta(\omega_k-\mu)}\right]+
\log\left[ 1+e^{-\beta(\omega_k+\mu)}\right]\right\}
\label{ferfreeparfun}
\end{equation}

The net free fermion charge density is 
\begin{equation}
\frac{1}{\beta L}
\frac{\partial}{\partial\mu}\log Z_F(T,\mu,L)=-2 F(T,\mu,L)
\label{rhofree}
\end{equation}
 $F$ being given in (\ref{FLfin}). Thus, from (\ref{rhofree}) 
and (\ref{parfunfin}) we have
\begin{equation}
Z(T,\mu)= Z(T,\mu=0,L)\exp \left\{-2\beta L\left[ \pi F^2 (T,\mu,L)+
\int_0^\mu d\mu'F(T,\mu',L)\right]\right\}
\label{parfunfin2}
\end{equation}
which is our  final result for the partition function at $\mu\neq 0$ for 
finite $L$. 
Now, let us analyse the behaviour of the function $F$ in the limit 
$L\rightarrow\infty$. In such a limit, the sum over $k$  becomes a trivial 
integral, which yields:
\begin{equation}
F(T,\mu,L\rightarrow\infty)=-\frac{\mu}{2\pi}
\label{FLinf}
\end{equation}

%\begin{equation}
%Z_F (T,\mu,L)=Z_F(T,\mu=0,L)\exp [-2\beta L\int_0^\mu d\mu'F(T,\mu',L)]
%\label{ferfreeparfunfac}
%\end{equation}

Thus, in the $L\rightarrow\infty$ 
limit, the $\mu$ dependence in (\ref{parfunfin2}) 
exactly cancels out. We get for the full partition  
function:
\begin{eqnarray}
Z(T,\mu,L\rightarrow\infty)&=& Z(T,\mu=0,L\rightarrow\infty)=
Z_{EM}Z_F(T,\mu=0,L\rightarrow\infty)\nonumber\\
&\times& 
\exp\left[ \frac{1}{2}\Tr\log (1+\Pi D)^{-1}\right]
\label{parfunLinf}
\end{eqnarray}
 which is only $T$ dependent. 
Its explicit expression can be found in section
IV of \cite{rual87}. 
That is, in the $L\rightarrow\infty$ 
limit the system bosonizes as well, and 
the only effective degrees of freedom are those of a massive 
boson field,  the net fermionic charge of the system being 
$\rho=(\beta L)^{-1}\partial \log Z /\partial\mu=0$. We 
remark that: 
i) This is a nonperturbative 
effect (which has been established due to the fact
that the Schwinger model at finite chemical potential and temperature 
 can, still, be solved exactly);  ii) It is 
characteristic of two dimensions. Recall for instance that 
 in perturbative four-dimensional QED in the infinite volume limit, 
 the free energy $\log Z$ is $\mu$-dependent and hence $\rho\neq 0$ 
 \cite{mor78};   iii) The 
 result (\ref{parfunLinf})  holds in the $L\rightarrow\infty$ limit: 
if we keep $L$ finite, $F$ is no longer given by 
 (\ref{FLinf}) but it acquires further corrections and the  study of the
counterpart of (\ref{parfunLinf})
will not be attempted here.

We now turn to the issue of the $C$-violating phase factor,  
 addressed previously in section 
 \ref{genfunden}. Thus, we consider the fermionic generating functional 
$Z_f[A,\xi,\overline\xi]$, treating now $A$ as an 
external background field.  
We  apply (\ref{funmetthird}) and we arrive at an 
 expression analogous to (\ref{genfunfried}), but now taking 
 $J=0$, omitting the derivatives with respect to $A$ and replacing 
$e\rightarrow -ie$. By recalling the regularised fermion closed 
 loop functional obtained in (\ref{Lreg}), we get:
\begin{eqnarray}
Z_f[A,\xi,\overline\xi]&=&Z_F Z_{EM}
\exp\left[-i\intt d^2 y \overline\xi (x) 
G(x,y,eA;\mu)\xi (y)\right]\nonumber\\
&\times&\exp\left[-\frac{1}{2}
\intt d^2 y A_\alpha (x) \Pi^{\alpha\beta}(x-y)
A_\beta (y)\right]\nonumber\\
&\times& \exp\left[-2ieF(T,\mu,L)\intt A_0(x)\right]
\label{ferfunfmet}
\end{eqnarray}

The comparison with 
(\ref{zfsepzero}) in the trivial sector leads to identify: 
\begin{equation}
{\cal J}^{(0)} (A,\mu)=-2eF(T,\mu,L)\intt A_0(x)
\label{jident}
\end{equation}
that has the form 
that we had anticipated in (\ref{jterm}), based upon the $C$ 
symmetry properties  of the phase of the fermionic determinant, and, so,
$\tilde F=-2eF$.

%\begin{equation}
%\frac{\partial}{\partial\mu}\log\det H(\tilde A;\mu)=\beta\sum_k
%\sum_{\stackrel{s_1=\pm}{s_2=\pm}} 
%\frac{s_1}{1+e^{\beta\left(\omega_k-\bar h_1-s_1 \mu+is_2\bar h_0\right)}}
%\label{derdethmu}
%\end{equation} 

We shall now provide an interesting check of consistency. On the basis of 
(\ref{parfunfor}), (\ref{facinstatriv}), 
(\ref{parfunh}) and (\ref{phaseh0}), we can calculate the partition 
function 
 through the method developed in section \ref{genfunden}. We shall do so 
in the $L\rightarrow\infty$ limit. 
Let us first differentiate with respect to $\mu$ in 
(\ref{parfunh}) and then take the $L\rightarrow\infty$ limit by replacing 
$\omega_k$ by continuous $\omega$ and 
 $\sum_k$ by  $(L/2\pi)\int d\omega$. 
The resulting integral can be done using 
\begin{equation}
\int_{-\infty}^{+\infty}d\omega \left[\frac{1}{1+e^{\beta(\omega-a+b)}}-
\frac{1}{1+e^{\beta(\omega-a-b)}}\right]=-2b
\label{intrule}
\end{equation}
which  is convergent when both pieces of the 
integrand are added together. Then,  in the  $L\rightarrow\infty$ 
limit we 
obtain: 
\begin{equation}
\frac{\partial}{\partial\mu}
\log\det H(\tilde A;\mu)=\frac{2\beta L\mu}{\pi}
\end{equation}
and, hence:
\begin{equation}
\sqrt{\det H}(\tilde A;\mu,T)=\sqrt{\det H}(\tilde A;\mu=0,T)\exp 
\left(L\beta\frac{\mu^2}{2\pi}\right)
\label{mudepdeth}
\end{equation}

We now insert the above result in (\ref{facinstatriv}) and, then,
 in (\ref{zfsepzero}), for the trivial sector, and  we compare to 
 (\ref{ferfunfmet}). We see that, before integrating out the 
photon field, 
 the results obtained with both methods are consistent with each other 
 in $L\rightarrow\infty$,   
since we get the same 
$\mu$ dependence, namely, the one coming from  the phase 
 factor (the explicit form of which  has been obtained with  
the second method) times that of the free 
 fermionic partition function, as it appears in (\ref{parfunfin2}) 
 when $F\rightarrow -\mu/2\pi$. To complete the check, we 
shall see  that the integration over the gauge 
boson fields also gives the same result with both methods in 
the $L\rightarrow\infty$ limit, namely, that in (\ref{parfunLinf}). 
From our previous discussion, it follows that the only part that 
depends on $\mu$ in the full partition function is the integral 
over the $(h_0,h_1)$ fields, the integrand of which is the exponential 
of  (\ref{phaseh0}) for $\Phi=0$ times $(-2ieF)$, multiplied by
$\sqrt{\det H(\tilde A)}$. From (\ref{mudepdeth}), it follows 
that all the 
$h$ dependence of $\det H$ is contained in the $\mu=0$ part. 
In order to extract the dependence on  $h_0$, we differentiate again 
  in (\ref{parfunh}). Using again (\ref{intrule}) we get in the 
$L\rightarrow\infty$ limit:
\begin{eqnarray}
\frac{\partial}{\partial h_0}\log\det H(\tilde A;\mu=0;h_0,h_1)&=&
2ie\sum_{s=\pm}\int_{-\infty}^{+\infty} d\omega \frac{s}
{1+e^{\beta\left(\omega-\bar h_1-is\bar h_0\right)}}\nonumber\\
&=&-\frac{8\pi L e^2 h_0}{\beta}
\label{derdethh0}
\end{eqnarray}
 which turns out to be independent on $h_1$. Thus, from 
(\ref{mudepdeth}) and
(\ref{derdethh0}), we derive:
\begin{eqnarray}
\sqrt{\det H}(\tilde A;\mu,T;h_0,h_1)&=&\sqrt{\det H}(\tilde A;
\mu=0,T;0,h_1)\nonumber\\
&\times &\exp \left(L\beta\frac{\mu^2}{2\pi}-\frac{2\pi Le^2 h_0^2}
{\beta}\right)
\label{depdeth0}
\end{eqnarray}

At this stage, we have to integrate over the fields 
 $(h_0,h_1)$. As  commented at the beginning of section \ref{genfuntri}, 
it is necessary to 
integrate over $h_0\in\IR$, 
to achieve consistency. This is reinforced by the 
topological term (\ref{jident}) in 
the phase of the fermionic generating functional. 
  Thus, from (\ref{phaseh0}) and (\ref{depdeth0}), it follows
that the relevant factor carrying the $\mu$ dependence is:
\begin{equation}
e\int_{-\infty}^{+\infty} dh_0 
\exp\left(-2i\mu h_0 L-\frac{2\pi L h_0^2}
{\beta}\right)=
e\sqrt{\frac{\beta}{2L}}\exp\left(-\frac{\mu^2 L\beta}{2\pi}\right)
\label{corrh0}
\end{equation}    
It turns out that the exponential 
$\mu$ dependence in (\ref{corrh0}) cancels 
that in (\ref{depdeth0}), and, hence, we arrive, again 
at the result (\ref{parfunLinf}), obtained with our previous method. 

\subsection{ The boson  propagator, the screening length and 
 the Polyakov loop }
 The exact boson 
propagator, that results after the integration of both fermion 
and gauge
fields,
can be obtained, 
in the trivial sector,  
 just 
by differentiating (\ref{genfunJ})  
with respect to  the external sources and 
setting $J=0$. We find:
\begin{equation}
\ID_{\alpha\beta}^{(\mu)} (x-y)=\ID_{\alpha\beta}^{(0)} 
(x-y)-(2eF)^2\int_T 
d^2 u d^2 z \ID_{\alpha 0}^{(0)} (x-u) 
\ID_{\beta 0}^{(0)} (y-z)
\label{bosproppos}
\end{equation}
 $\ID_{\alpha\beta}^{(0)} (x-y)$ being given in (\ref{ddprop}). 
The above 
expression is, again,  formal, in the sense 
that we have to specify 
the order in which the spatial and temporal integrals 
should be done in 
the second piece on 
the right-hand-side. If we adopt the same prescription as
 that in 
 section \ref{charparfun} 
(that is, by doing the integrals in the order that 
gives a finite 
 answer) we arrive in momentum space at:
\begin{eqnarray}
\ID_{00}^{(\mu)}(\omega_n,\omega_k) &=&\ID_{00}^{(0)} 
(\omega_n,\omega_k)-\delta_{n0}\delta_{k0}\beta L
\frac{4\pi^4 F^2(T,\mu,L)}{e^2}\nonumber\\
\ID_{0i}^{(\mu)}(\omega_n,\omega_k) &=&\ID_{i0}^{(\mu)}
(\omega_n,\omega_k)=\ID_{0i}^{(0)} (\omega_n,\omega_k)
\nonumber\\
\ID_{ij}^{(\mu)}(\omega_n,\omega_k) &=&\ID_{ij}^{(0)} (\omega_n,\omega_k)
\label{bospropmom}
\end{eqnarray} 
$\ID_{00}^{(0)} (\omega_n,\omega_k)$ being given by the expression 
between  
square brackets in (\ref{ddprop}). In the 
 $L\rightarrow\infty$ limit, $\omega_k$ is replaced by a continuous 
variable $\omega$
 and $L\delta_{k0}F^2$ goes to  $(\mu^2/4\pi^2)\delta (\omega)$, in  
(\ref{bospropmom}). 
 As a check of consistency, we notice 
 that the propagator (\ref{bospropmom}) (which is $\mu$ dependent) 
satisfies the gauge invariance 
condition $\omega^\mu\omega^\nu \ID_{\mu\nu}=\alpha$, with 
 $\alpha$ the gauge-fixing parameter.  
For finite $L$, the only 
 Euclidean  pole of the propagator (\ref{bospropmom}) 
 remains at $\omega^2=-m^2=-e^2/\pi$, which defines the mass of 
the boson and is independent on $\mu$. However, in the 
 $L\rightarrow\infty$ limit, a new pole appears at $\omega^2=0$ 
due to the contribution of $\delta (\omega)$ in the 
second ($\mu$ dependent) term. To clarify what this pole means 
physically, 
let us calculate first 
 the inverse correlation 
length squared $M^2$  in the  $L\rightarrow\infty$ limit, 
 through  the usual definition 
\begin{equation}
M^2= {\cal D}_{00} (\omega_n=0,\omega\rightarrow 0)
\label{screendef}
\end{equation}
where  ${\cal D}_{\alpha\beta} (\omega_n,\omega)$ is the  boson 
self-energy, which is defined in momentum space through
\begin{equation}
{\cal D}_{\alpha\beta}(\omega_n,\omega)=
\left[\ID ^{-1}\right]_{\alpha\beta}^{(\mu)}
(\omega_n,\omega)-\left[ D^{-1}\right]_{\alpha\beta}(\omega_n,\omega)
\end{equation}
 Accordingly, from (\ref{screendef}) 
and the propagators in  (\ref{freeprop}) 
and  (\ref{bospropmom}), we obtain 
\begin{equation}
M^2(T,\mu)=0
\label{msqu}
\end{equation} 

That is, the  mass $M^2$ vanishes in the  $L\rightarrow\infty$ limit.
 This is indeed a consistent result if we 
 recall the connection between the screening length and the 
 equation of state of the system given in \cite{kap89}:
\begin{equation}
M^2(T,\mu)=e^2\frac{\partial}{\partial\mu} \rho (T,\mu)
\label{screenes}
\end{equation}
 $\rho$ being the total fermion  charge  density. Hence (\ref{msqu}) 
and  (\ref{screenes})  are 
 consistent  in the $L\rightarrow\infty$ limit:
 we got a total zero  charge 
density, while 
  $M^2$ tends to zero. However, were  $M^2$  be interpreted as a
  vanishing screening 
mass, then  the system would be in a confined phase and 
 the $Z$ 
 symmetry would be  restored, 
which does not occur for $\mu=0$ and $T\neq 0$  in the 
 $L\rightarrow\infty$ limit \cite{griso96}. 
 We shall
outline below a correct understanding of the screening mass. 
 The order parameter of this 
 symmetry is the thermal average of the Polyakov loop
\begin{equation}
 P_{\tilde e} (x^1)\equiv \exp\left[ i\tilde e\int_0^\beta 
 d\tau A_0(\tau,x^1)\right]
\end{equation}
for an arbitrary charge $\tilde e$ and a spatial point $x^1$. 
In addition, 
 the correlator of two Polyakov loops with charges $\tilde e$ and 
 $-\tilde e$ and spatial points $x^1$ and $u^1$ measures, at 
large spatial 
 separation,  the effective potential 
 between both charges. Thus,  in order to clarify  the role of the two 
 different masses $m$ and $M$, 
let us calculate the Polyakov loop and the correlator 
 for $\mu\neq 0$ in the $L\rightarrow\infty$ limit. 
 First, we shall use the functional methods developed in section 
 \ref{genfuntri}.  
 After the obvious 
relabelling of variables,the insertion of $P_{\tilde e} (x^1)$ in 
 the functional integral (\ref{genfun}) amounts formally to replace 
 $J(y)\rightarrow iP(y)$ with $P_0(y)=\tilde e\delta (y^1-x^1)$ and 
 $P_1(y)=0$. Then, recalling our result (\ref{genfunJ}), we get

\begin{eqnarray}
\langle  P_{\tilde e} (x^1)\rangle =  \exp\left[-
\frac{1}{2}\int d^2 y d^2 z  \ID_{00}^{(0)} (y-z) P_0(y)
\left(P_0(z)-2G_0(z)\right)\right]
=\exp\left[-\frac{\beta{\tilde e}^2}{2m}-\beta\mu\frac{\tilde e}{e}\right]
\label{polymu}
\end{eqnarray}
where in the last 
step  we have first followed our  convention of performing  
 the time integration before the spatial one and then we have taken 
 the limit $L\rightarrow\infty$, so as to transform 
 spatial sums into 
integrals. The result  (\ref{polymu}) can also be obtained 
 using the 
decomposition (\ref{deco1}) and (\ref{deco2}) for the gauge field. 
  As  we saw in previous sections, all the $\mu$ dependence is included 
 in the $h_0$-dependent part. Then, since 

\begin{equation}
\int_0^\beta d\tau A_0 (\tau,x^1)=2\pi h_0-\partial_x^1\int_0^\beta d\tau 
 \phi(\tau,x^1),
\end{equation}
all we have to do 
is to insert a piece $\exp (2\pi i h_0 {\tilde e}/e)$ in the 
 integrand in (\ref{corrh0}), in the $L\rightarrow\infty$ limit. 
 It is straightforward to  
  get again the $\mu$ dependence of the Polyakov loop in (\ref{polymu})
consistently, 
 once we integrate over $h_0\in\IR$.

 We see that $\langle  P_{\tilde e} (x^1)\rangle$ 
never vanishes for any value of $T$ and 
 $\mu$, so that the $Z$ symmetry is never restored for 
$L\rightarrow\infty$.  Our result extends that obtained 
in \cite{griso96} for $\mu=0$. Then, there 
 should exist a 
nonvanishing screening mass, which, by consistency with our 
 result for the propagator, 
should be  $m=e/\sqrt{\pi}$, independent on $\mu$. 
 This is confirmed by calculating the correlator of two Polyakov loops. 
 Following the same steps as  for $\langle P\rangle$ we obtain with 
 both methods that 
$\langle  P_{\tilde e} (x^1)   P_{-\tilde e} (u^1)\rangle$ 
 is independent on $\mu$ and 

\begin{equation}
\lim_{x^1-u^1\rightarrow\infty}
\langle  P_{\tilde e} (x^1)   P_{-\tilde e} (u^1)\rangle
=\exp\left[
-\frac{\beta {\tilde e}^2}{2m}\right],
\end{equation}
 that is,  $m$ is 
the screening mass between the two opposite charges. It 
 seems  clear that the definition (\ref{screendef}), leading to 
 (\ref{msqu}), does not 
give the right answer for the screening mass between 
 two opposite charges, nor can be used to infer any conclusion about the 
 breaking of the confinement or $Z$ symmetry. However, it is 
 the  consistent equation to  use 
if we want to get  the equation of 
 state of the system through (\ref{screenes}).

Notice that we have restricted ourselves to the trivial
sector. The boson propagator could receive additional contributions
 from sectors with $k\geq 1$, apart from that in (\ref{bosproppos}). 
 However, we have seen 
that $\rho$ only receives contributions from the trivial
sector, so that,  (\ref{msqu}) is valid for any sector 
if (\ref{screenes}) holds. In addition, it is enough to restrict 
ourselves to the trivial sector to calculate  
correlators   of Polyakov loops \cite{griso96}. Hence, we expect 
 the above results 
for the screening mass to remain valid for  $\Phi\neq 0$. 

If the length of the system $L$ is kept finite, the 
 definition (\ref{screendef}) for $M^2$ is no longer valid. Using 
(\ref{screenes}) instead  would 
 give rise to a nonzero and  $\mu$-dependent $M^2$, directly from 
(\ref{parfunfin2}).  Remember that for 
 finite $L$ there is no pole of the propagator at $\omega^2=0$. 
 However, it is not clear  whether 
 (\ref{screenes}) remains valid for finite $L$. On the other hand, 
 for finite $L$ and $\mu= 0$, the screening mass is only different 
 from zero if $\tilde e/e\in\IZ$ \cite{griso96}.  
 A more rigorous analysis of the finite length corrections for 
 $\mu\neq 0$ is beyond 
the scope of this work.

\section{The Thirring model at finite $T$ and $\mu$}
\label{thirring}

\subsection{The generating functional}

We shall consider a system of many massless fermions (and antifermions) 
in one dimension 
 (inside a finite interval of length $L$) at equilibrium at absolute 
temperature $T$ and chemical 
 potential $\mu$. By assumption, their dynamics is described by the 
Thirring quartic lagrangian. 
 Let $\xi$,  $\overline\xi$ be fermionic external sources. Then, 
in the imaginary time formalism, 
 the generating functional of the system reads  now 
\begin{eqnarray}
Z[\xi,\overline\xi]&=& N(\beta,\mu)\int_{antiperiodic} 
\!\!\!\!\!\!\!\!\!\!\!\!\!\!\!\!\!\!\!
{\cal D}\overline\psi {\cal D}\psi
 \exp \left[\intt \left(-\overline\psi (\pabar -\mu\gamma^0) \psi
+\overline\xi\psi+\overline\psi\xi\right)\right.\nonumber\\
 &-&\left[\frac{g^2}{2}\left(\overline\psi\gamma^\nu\psi\right)
\left(\overline\psi\gamma_\nu\psi\right)\right]
\label{thirringgenfun}
\end{eqnarray}
where $g$ is the coupling constant. The partition function 
is $Z[0,0]$. One can also cast $Z[\xi,\overline\xi]$ as follows
\begin{eqnarray}
Z[\xi,\overline\xi]&=&\exp
\left[-g\intt \frac{\delta}{\delta\xi (x)}\gamma_\nu 
\frac{\delta}{\delta J_\nu (x)}\frac{\delta}{\delta\overline\xi (x)}
\right]\nonumber\\
&\times&
 Z_1 [\xi,\overline\xi;J]\vert_{J=0}\label{thirringz}\\
Z_1 [\xi,\overline\xi;J]&=&N Z_F \exp\left[\frac{1}{2}\int_T d^2 x' d^2 y' 
J_\alpha (x') K^{\alpha\beta} (x'-y') J_\beta (y')\right]\nonumber\\
&\times&
\exp \left[\int_T d^2 x'' d^2 y'' \overline\xi 
(x'')S(x'',y'';\mu)\xi (y'')
\right]
\label{thirringz1}\\
K^{\alpha\beta} (x'-y')&=&\frac{1}{\beta L}
\sum_{n,k=-\infty}^{+\infty}e^{i
\omega\cdot 
(x'-y')}\left[\delta^{\alpha\beta}-f(\omega^2)\frac{\omega^\alpha 
\omega^\beta}{\omega^2}\right] 
\label{thirringk}
\end{eqnarray}
with  $S$ in (\ref{freeprop}),  $Z_F$ the free fermionic 
partition function
 and $J=(J_0,J_1)$  a boson source, to be set equal to zero after all 
functional 
 differentiations with 
respect to it have been performed in (\ref{thirringz})
 and  
  $\omega$ in (\ref{thirringk}) is the same as in the gauge boson free 
propagator in
  (\ref{freeprop}). We have introduced an arbitrary   
function $f(\omega ^2)$, by virtue of the fact that the 
current $\overline\psi\gamma^\nu \psi$ in the actual Thirring model 
is conserved. A proof of 
 equations (\ref{thirringz}) and (\ref{thirringz1}) 
follows readily 
through  steps similar to those in \cite{rual87,fried72}. 
At this stage, using standard techniques \cite{fried72}, 
 one can rewrite eqs.(\ref{thirringz}) and (\ref{thirringz1})  as 
\begin{eqnarray}
Z[\xi,\overline\xi]&=&\exp\left[-\frac{1}{2}\int_T d^2 x' d^2 y' 
\frac{\delta}{\delta A^*_\alpha (x')} K^{\alpha\beta} (x'-y')
 \frac{\delta}{\delta A^*_\beta (y')}\right] \nonumber\\
&\times& Z_f [A^*_\nu,\xi,\overline\xi]
\vert_{A^*=0}\label{thizzf}\\
 Z_f [A^*_\nu,\xi,\overline\xi]&=&N Z_F\exp\left[-ig\intt 
\frac{\delta}{\delta\xi (x)}
 A^*_\nu (x) \frac{\delta}{\delta\overline\xi (x)}\right]\nonumber\\
&\times &\exp\left[
\int_T d^2 x'' d^2 y'' \overline\xi (x'') S(x'',y'';\mu)\xi (y'')\right]
\label{thizf}\\
A^*_\nu (x)&=&-i\int_T d^2 z K_{\nu\beta} (x-z) J^\beta (z)
\label{thia*}
\end{eqnarray}

Again, one sets $A^*=0$ in the above equations, after having carried out 
all functional 
differentiations.
   Standard 
functional techniques allow us now to establish that $Z_f [A^*_\nu,\xi,
\overline\xi]$, as 
given in (\ref{thizf}) also coincides with the right-hand-side of 
(\ref{fergenfun}) (when 
 due care is taken of the normalisation factor $N(\beta,\mu)$) 
provided that, in the latter, 
one replaces $e$ by $g$ and $A$ by $A^*$. Let us 
concentrate on $A^*$ belonging to the trivial sector 
(see comments below). 
Then, 
 by recalling the developments in section 
 \ref{genfuntri}, one finds immediately
\begin{equation}
Z_f [A^*_\nu,\xi,\overline\xi]=Z_F \exp\left[-i\intt\int_T d^2 y 
\overline\xi (x) G(x,y,gA^*;\mu)\xi (y) +L[A^*]\right]
\label{thisch}
\end{equation}
where  $G(x,y,gA^*;\mu)$ 
and $L[A^*]$ are now given by the right-hand-sides of 
(\ref{gansatz}), (\ref{solchi}) and (\ref{piexp}) (with the same $S$, 
 $\Delta$, $\Pi$ and $F$), when one 
replaces $e,A$ by $g,A^*$, respectively. Thus, we have provided 
the solution
 for the Thirring 
 model at finite $T$, $\mu$ in terms of the fermionic generating 
functional 
for the 
Schwinger model, without zero modes ($\Phi =0$). The formal use  of 
eqs.(\ref{thirringz})-(\ref{thirringk}) and (\ref{thizzf})-(\ref{thia*}) 
(which can be checked   upon comparing  the corresponding  
expansions into 
powers of $g$) 
 is justified if we restrict  $Z_f[A^*_\nu,\xi,\overline\xi]$ to the 
trivial sector of the Schwinger 
 model, as implemented through (\ref{thisch}). 
 In this regard, it is 
 interesting to note that, upon using  (\ref{thia*}) 
in (\ref{axanom}) and 
taking into 
account that  the propagator $K_{\alpha\beta}$ in (\ref{thirringk}) 
satisfies periodic boundary 
 conditions in both $x^0$ and $x^1$ directions, one immediately  
finds 
$\Phi[A^*]=0$, which  confirms that
 $Z_f[A^*,\xi,\overline\xi]$ should be restricted to the 
trivial sector 
 when use is made of (\ref{thizzf})-(\ref{thia*}) 
and, hence, the consistency of 
 (\ref{thisch}). On the other hand, this appears also 
to be consistent with the idea that, in the end, we 
are going to set 
$A^*=0$ and then 
 we can take  the vector field $A^*_\nu$  as a configuration 
in the 
trivial sector, 
 that is, topologically connected with $A^*=0$. 
 It is unknown to us whether the Thirring  model 
may have other solutions 
(besides that 
 given in (\ref{thizzf}) and (\ref{thisch})).

\subsection{The thermodynamical partition function and fermion 
correlation function}

The thermodynamical partition function becomes, upon applying  
(\ref{recetafried3}) to 
 (\ref{thizzf}) and (\ref{thisch})

\begin{eqnarray}
Z[0,0]=Z_F \exp\left\{\frac{1}{2}\Tr\log 
(1+\Pi K)^{-1}\right.\nonumber\\
-\left.\frac{1}{2} (2gF(T,\mu,L))^2\intt d^2y
\left[K(1+\Pi K)^{-1}\right]^{00} 
  (x-y)\right\}
\label{thifunpaLfin}
\end{eqnarray}
where in momentum space we have
\begin{eqnarray}
 \left[(1+\Pi K)^{-1}\right]^{\alpha\beta}(x-y)&=&\frac{1}{\beta L}
\sum_{n,k=-\infty}^{+\infty}e^{i\omega\cdot (x-y)}
  \left[\delta^{\alpha\beta}-\frac{g^2/\pi}{1+g^2/\pi}
\left(\delta^{\alpha\beta}-\frac{\omega^\alpha\omega^\beta}{\omega^2}
\right)\right]\nonumber\\
 \left[K(1+\Pi K)^{-1}\right]^{\alpha\beta}(x-y)&=&\frac{1}{\beta L}
\sum_{n,k=-\infty}^{+\infty}e^{i\omega\cdot (x-y)} 
 \left[\frac{\delta^{\alpha\beta}}{1+g^2/\pi}\right.\nonumber\\
&-&\left.\frac{\omega^\alpha\omega^\beta}{\omega^2}\left( f(\omega^2)-
\frac{g^2/\pi}{1+g^2/\pi}\right)\right]
\end{eqnarray}

 Using (\ref{FLinf}), (\ref{thifunpaLfin}) yields readily 
for $L\rightarrow\infty$
\begin{eqnarray}
Z[0,0]&=&Z_F(T,\mu)\exp\left\{ La\log\frac{1}{1+g^2/\pi}
\right\}\exp\left\{-\beta L\frac{g^2\mu^2}{2\pi^2}
\left[ \frac{1}{1+g^2/\pi}\right.\right.\nonumber\\
&-&\left.\left.b\left(f(0)-\frac{g^2/\pi}{1+g^2/\pi}\right)\right]\right\}
\label{thifunpaLinf}
\end{eqnarray}
where
\begin{eqnarray}
a&=&\frac{1}{2}\sum_{n=-\infty}^{+\infty}\int_{-\infty}^{+\infty}
\frac{dk}{2\pi}\nonumber\\
b&=&\lim_{L\rightarrow\infty}
\frac{1}{(\beta L)^2}\intt d^2 y\sum_{n,k=-\infty}^{+\infty}
e^{i\omega\cdot (x-y)}\frac{\omega^0\omega^0}
{\omega^2}
\label{ab}
\end{eqnarray}

Notice that the first exponential on the right-hand-side 
of (\ref{thifunpaLinf}) is independent on $\beta$, $\mu$ and then 
it is irrelevant as far as the thermodynamics of the model is concerned. 
On the other hand, the constant $b$ (which is 
 independent on both $T$ and $\mu$) could in principle give rise to a 
dependence on $f(0)$. Again, the very definition of $b$ is 
 ambiguous. If we agree to evaluate the (imaginary time) integrals over 
$x^0$ and $y^0$ in (\ref{ab}) before the spatial ones, then 
 $b=0$ and, hence, the independence of the partition function on 
$f(\omega^2)$ follows, which is a welcomed result. Another 
 independent reason to favour this prescription is that it is the 
same as that leading from (\ref{parfundd}) to (\ref{parfunfin}). 
 Now, by recalling the expression for the free charge density in 
(\ref{rhofree}), we have from (\ref{thifunpaLinf})
 that the total fermion number density of the  system reads in the 
$L\rightarrow\infty$ limit
\begin{equation}
\rho=\frac{\mu}{\pi+g^2}
\label{thirho}
\end{equation}

Hence we obtain that the Thirring model at finite $T$ and $\mu$ is 
no longer a free fermion gas, but the fermion density 
acquires a correction in $g^2$, as it stands in (\ref{thirho}).
 It differs from the result in \cite{yoko87} in which only the free 
contribution to the fermion density 
 remains. It is 
 clear from our analysis starting from the Schwinger model that
 the correction to the free gas comes entirely from the topological 
contribution depending on the $F$ function.  
 This contribution only depends on the harmonic field 
 $h_0$ in the decomposition of the gauge field. The calculation in 
\cite{yoko87} was done in 
real time formalism, in which      this term is not present, whereas in 
\cite{sawi96} 
the model is solved in the torus. 
As it is emphasised in \cite{sawi92,sb9495,sawi96}, the toroidal 
 compactification  is very useful 
 to deal with infrared divergences and the harmonic parts 
of the gauge field are essential to correctly quantise the model. 
It  is  the most natural choice when using  
the imaginary time formalism, as we have done in this work. 
On the other hand, if we evaluate 
 the pressure of the system, which 
 follows directly by taking the logarithm of the partition function in 
(\ref{thifunpaLinf}), 
  our  result (not quoted for
brevity) agrees with \cite{sawi96}, which provides a check of 
consistency between our  
methods and those used in that work.

Finally, we shall give the exact fermion correlation function for 
the Thirring model at nonzero $T$ and $\mu$ 
\begin{eqnarray}
 G(x,y)&=&\left.\frac{\delta^2 
\log Z[\xi,\overline\xi]}{\delta\overline\xi(x)\delta\xi 
(y)}\right\vert_{\xi=\overline\xi=0} =\Theta (x,y) S(x,y)\nonumber\\
\Theta (x,y)&=&\exp\left\{- g^2\frac{1}{\beta L}
\sum_{n,k=-\infty}^{+\infty}\frac{1}{\omega^2}
\left[1-e^{i\omega\cdot (x-y)}\right] 
\right.\nonumber\\
&\times&\left.
 \left[ f(\omega^2)-\frac{g^2/\pi}{1+g^2/\pi}
\right]+2gF(T,\mu,L)\cdot c\right\}
\end{eqnarray}
  where we have used  again (\ref{recetafried3}) into 
(\ref{thizzf}) and (\ref{thisch}) and performed 
the functional differentiation. In turn, $c$ is given by 
the formal expression 
\begin{equation}
c=\sum_{n,k=-\infty}^{+\infty}\left[e^{-i\omega x}-e^{-i\omega y}\right]
\frac{\delta_{0,n}\delta_{0,k}}{\omega^2}\left[
-\frac{\omegabar\gamma^0}{1+g^2/\pi}+\omega_0
\left( f(\omega^2)-\frac{g^2/\pi}{1+g^2/\pi}\right)\right]
\label{c}
\end{equation}
which, again, turns out to be ambiguous. Like we did with 
the same ambiguities before, let us evaluate the summation over $n$ in 
  (\ref{c}) (which is reminiscent of imaginary-time integrations) 
before the spatial summation and let $L\rightarrow\infty$. Then 
 one gets 
\begin{equation}
c=\frac{i}{1+g^2/\pi}(x_1-y_1)\gamma^1\gamma^0
\end{equation}

\section{Conclusions and discussion}
\label{conc}

The main new results obtained in this work are the following:

1) In the imaginary 
time formalism, the fermionic generating functional $Z_f$ 
with an external electromagnetic 
 field and the full generating functional $Z$ for the Schwinger model 
have been explicitly obtained for any 
 spatial length $L$ 
 in the trivial sector (in which the Dirac operator has no zero modes). 

2) The work previously done in \cite{sawi92} at finite $T$ 
but $\mu=0$, 
  in which the model was formulated in a two-dimensional torus for 
an arbitrary
 number of 
 zero modes, can be extended when both $T$ and $\mu$ are nonzero. 
  Such an extension has to be worked out carefully 
 due to  some non trivial peculiarities of the 
$\mu\neq 0$ case. Technically, the main distinctive feature is the 
lack of hermiticity of the Dirac operator. This implies  a non-vanishing 
phase factor ${\cal J}^{(k)}$
for the fermion determinant in the sector with $k$ zero modes. 
  Using functional methods 
we have evaluated this term for $k=0$ (the trivial
sector), which plays 
  a crucial role in the solutions 
 for the Schwinger and Thirring models presented here and in the physical 
features thereof. That phase depends on $T$ and $\mu$, 
 is linear in the zeroth component of the electromagnetic 
potential $A_0$ (in agreement with charge conjugation symmetry 
 arguments) and  vanishes if $\mu=0$ for any $T,A_0$. 
 Furthermore, this term is topological, in the sense that it  
changes only under nontrivial gauge transformations,  with nonzero 
winding number around the 
 $S^1$ parametrising the Euclidean time. In terms of the 
Hodge decomposition of the gauge field in the torus, it only  depends
   on the harmonic part. The existence of  topological $\mu$-induced 
 effective actions     
 seems to be a common feature of different  models  
\cite{rewi85,varios8595}. 

3) For the Schwinger model  we have calculated  the thermodynamical 
partition function. The topological  phase factor in the 
 effective action gives rise  to a nonperturbative contribution that 
in the $L\rightarrow\infty$ limit 
 exactly cancels the $\mu$ dependence contained in the  free fermionic 
partition function. Then, at  $L\rightarrow\infty$, the 
 partition function is independent on $\mu$ and hence the total  
charge density of the system is zero. In other words, the 
 system  bosonizes even though  it could have a net fermionic 
charge density at nonzero $\mu$.
   The partition function factorises, as in \cite{rual87} (into 
 that of free 
fermions 
 at $\mu=0$ times a factor, which is the ratio of that of  
massive bosons, divided by
 that of free massless bosons), the mass of the boson 
being $m=e/\sqrt{\pi}$, 
independent of $T$ and $\mu$. The exact boson propagator has 
an additional $\mu$-dependent piece for any $L$. At  
$L\rightarrow\infty$ this new 
 piece  gives rise to  a vanishing inverse correlation 
length squared $M^2$, 
 which is interpreted through 
the relationship between $M^2$ 
and the first derivative of the 
charge density given in   \cite{kap89}. However, by calculating
 for  $\mu\neq 0$
 the thermal average of the Polyakov loop (which is $\mu$-dependent)  
and  its correlator ($\mu$-independent), we have 
 shown that the $Z$ symmetry is broken for any $T$ and $\mu$ 
(deconfinement) and that 
 the  screening mass between two opposite charges 
is the mass $m$ of the  boson.  
  A study of what happens regarding the above mentioned 
 cancellation in the thermodynamical partition 
function, when $L$ is kept 
finite,  lies outside the scope
of this work. Our 
 computations of the thermodynamical partition function 
through two different 
methods yielding the same result establish 
 the consistency of our approach.  

4) Several important features of the solution of the Schwinger 
model for $\mu\neq 0$ 
in the sectors with zero modes are summarised,  as they are closely 
related to the analysis in the trivial sector. 
 Namely, we have given the general structure of the fermion 
determinant, solving the spectrum of the Dirac operator 
 for an instanton configuration when $\mu\neq 0$. The 
chemical potential breaks the chiral degeneracy of the spectrum. 
 We remark that the correlation functions for $\mu\neq 0$ have been 
analysed in \cite{chsch96}, although in that paper a different approach  
 based on bosonization is used  and the harmonic part of the gauge field 
(and hence the 
 contribution of the phase factors) is not considered. 
The analysis of the phase factors and the fermionic two point function 
when  there are zero modes lies beyond the scope of this work.
  
5) In the imaginary time formalism as well, the generating functional 
for the massless Thirring model at finite $T$, $\mu$ 
 is constructed in terms of the fermionic generating functional $Z_f$ 
for the Schwinger model, previously found in 
 this work. We have justified that it is enough to restrict ourselves 
to the 
trivial sector for $Z_f$. The thermodynamical 
 partition function, the total fermion number density and the fermion 
correlation 
function have been computed 
for non-vanishing $\mu$ and $T$. A distinctive 
 feature is that all of them depend nontrivially on $\mu$, as a 
consequence of
 the non-trivial phase ${\cal J}^{(0)}$ of the 
 Schwinger model. Our result for the pressure agrees with \cite{sawi96}, 
which shows that our 
different approach is consistent. Our total fermion density differs from
\cite{yoko87}, where it was obtained, using real time formalism, that the 
model is equivalent to that of free fermions. The 
origin of that discrepancy is that in \cite{yoko87} the harmonic pieces of 
the vector field are not considered.

\vspace{1cm}
{\Large \bf Acknowledgements}
\vspace{0.5cm}
 
This work was supported in part by the European Commission under the 
 Human Capital and Mobility programme, contract number ERB-CHRX-CT94-0423. 
The financial support of CICYT, project AEN96-1634 is also acknowledged. 
One of us (A.G.N.) has received financial support from Spanish Ministry 
of Education and Culture,  through a postdoctoral fellowship of the 
'Perfeccionamiento de  Doctores y Tecn\'ologos en el Extranjero' programme  
and he is very grateful to  Profs. T.Kibble, R.Rivers and 
T.Evans of the 
Theory Group at Imperial College, for  their kind  hospitality and for 
useful 
discussions and suggestions. We are also grateful to 
Prof.F. Ruiz Ruiz for providing some useful information.  

\appendix

\section{The fermion functional determinant for $\Phi\neq 0$}  

 We shall outline here the calculation of $\det ' H(A;\mu)$ in section  
\ref{instanton}, following  
 steps similar to those in \cite{sawi92}, with suitable generalisations 
for our present case. 
 Firstly, we shall relate $\det ' H(A;\mu)$ with 
$\det ' H(\tilde A;\mu)$. For that purpose, 
 we  define 
${\Dbar}_\alpha$ replacing $e\rightarrow e\alpha$ in (\ref{factalpha}).  
 Then, the corresponding $H_\alpha={\Dbar}_\alpha^\dagger {\Dbar}_\alpha$ 
interpolates between $H(A)$ and $H(\tilde A)$ when $\alpha$ varies 
from 0 to 1, and so on for 
$\overline H_\alpha={\Dbar}_\alpha {\Dbar}_\alpha^\dagger$. 
  The  operator $H_\alpha$  
 can be cast as in (\ref{h}), but now with $A_\mu=\tilde 
A_\mu-\alpha\epsilon_{\mu\nu}\partial_\nu \phi$
 and the operator $\overline H_\alpha$ is obtained from 
$H_\alpha$ by changing $\mu\rightarrow -\mu$.
 By using $\zeta$-
regularization \cite{ball89}, we have:
\begin{eqnarray}
{\log\det} ' H_\alpha&=&-\left.\frac{d}{ds}\zeta_H (s;\alpha)
\right\vert_{s=0}\nonumber\\
\zeta_H (s;\alpha)&=&\sum_{q=k+1}^{\infty}\mu_q^{-s} (\alpha)
\label{apzeta}
\end{eqnarray}
$\mu_q (\alpha)$ denoting, generically, the non-vanishing eigenvalues of 
$H_\alpha$.  As in section \ref{factzm}, we choose the eigenstates of 
 $\overline H_\alpha$ as  
$\varphi_q^{(\alpha)}=
({\Dbar}_\alpha\phi_q^{(\alpha)})/\sqrt{\mu_q(\alpha)}$, 
 where $\phi_q^{(\alpha)}$ are the eigenstates of $H_\alpha$  for 
$\mu_q (\alpha)\neq 0$. Now we  use the  Feynman-Hellmann 
 formula
 $\dot\mu_q (\alpha)=(\phi_q^{(\alpha)},\dot H_\alpha\phi_q^{(\alpha)})$,   
 where the dot indicates derivation 
with respect to $\alpha$,  
and the Seeley-de Witt expansion \cite{ball89} for 
$H_\alpha$. Then,     following  similar steps as in 
\cite{sawi92} we can write  the derivative of ${\log\det} ' H_\alpha$
 in (\ref{apzeta}) with respect to $\alpha$, in terms of   
$E_\alpha=\tilde E+\alpha\Delta\phi$, $\phi(x)$ and the zero modes 
 $\phi_p^{(\alpha)}$ and $\varphi_p^{(\alpha)}$ of 
 $H_\alpha$ and $\overline H_\alpha$. 
The latter are related to the zero modes
 of $H$ and $\overline H$ simply by multiplying by 
$\exp (-e\alpha\gamma_5)$. 
 Then, the integral in $\alpha$ can be done and we obtain

%\begin{equation}
%\dot\mu_q=2e\mu_q\left[(\phi_q,\gamma^5\phi,\phi_q)+
%(\varphi_q,\gamma^5\phi,\varphi_q)\right]
%\label{dotmuq}
%\end{equation}
%where $\varphi_q$ are the nonzero modes of 
%$\overline H_\alpha$, which, as in the 
%previous section, are chosen as 
%. Then, from (\ref{dotmuq}), 
%we have:
%\begin{equation}
%\frac{d}{d\alpha}\zeta_H(s;\alpha)=-\frac{2es}{\Gamma(s)}
%\int_0^\infty dt t^{s-1} \tr ' 
%\left[(e^{-tH_\alpha}+e^{-t\overline H_\alpha})\gamma^5\phi\right]
%\label{ddalphazeta}
%\end{equation}  
%where we have used a standard $\zeta$-function regularization 
\cite{ball89}. 
% The $\tr'$ symbol means that only the trace over the 
% nonzero modes is taken. Thus:
%\begin{equation}
%\tr ' (e^{-t\overline H_\alpha}\gamma^5\phi)=\intt x \phi(x)\tr_D 
%\gamma_5 \langle x\vert 
%e^{-t\overline H_\alpha}\vert x\rangle - \sum_{p=1}^k\intt 
%\phi^\dagger_p (x) \gamma_5 \phi \phi_p (x)
%\label{trsep}
%\end{equation}
% $\phi_p$ being the zero modes of $H_\alpha$. Now let us  
%introduce a mass parameter through $H_\alpha\rightarrow 
%H_\alpha + m^2$ and recall the Seeley-de Witt expansion 
\cite{ball89} for
%$H_\alpha$:
%\begin{equation}
%\langle x\vert e^{-t H_\alpha}\vert x\rangle=\frac{1}{4\pi t}
%(1+e E_\alpha \gamma^5 t+\Od (t^2)) 
%\label{sdwalpha}
%\end{equation}
%with $E_\alpha=\tilde E+\alpha\Delta\phi$. 
%Then, upon replacing (\ref{trsep}) and (\ref{sdwalpha}) in 
%(\ref{ddalphazeta}), 
%carrying out the Gaussian integrals  and 
% taking the derivative with respect to 
%$s$ in $s=0$, we get the following result (independent on $m^2$):
%\begin{equation}
%\frac{d}{d\alpha}\log{\det}' H_\alpha=\frac{2}{\pi} 
%e^2\intt \phi(x)E_\alpha (x)-e\intt \phi(x)\sum_{p=1}^k\left(
%\phi_p^\dagger\gamma^5 \phi_p+\varphi^\dagger_p\gamma^5 \varphi_p \right)
%\label{priortoint}
%\end{equation}
%  Now, let us recall that the zero modes of $H_\alpha$, 
%$\overline H_\alpha$ and those of $H$, $\overline H$ are related 
through: 
%\begin{eqnarray}
%\phi_p^{(\alpha)}&=&e^{-e\alpha\gamma^5\phi}\phi^{(\alpha=0)}\nonumber\\
%\varphi_p^{(\alpha)}&=&e^{-e\alpha\gamma^5\phi}\varphi^{(\alpha=0)}
%\label{relzemoalpha}
%\end{eqnarray}
%which implies that
%\begin{eqnarray}
%e\intt \phi(x)\sum_{p=1}^k \phi_p^\dagger\gamma^5 \phi_p=
%-\frac{1}{2}\frac{d}{d\alpha}\log\det N^{(\alpha)}\nonumber\\
%e\intt \phi(x) \sum_{p=1}^k \varphi^\dagger_p\gamma^5 \varphi_p=
%-\frac{1}{2}\frac{d}{d\alpha}\log\det M^{(\alpha)}
%\end{eqnarray}

\begin{eqnarray}
{\det}' H(A;\mu)&=&{\det}' H(\tilde A;\mu) \det\left\{ N^{(1)}
\left[N^{(0)}\right]^{-1}\right\} \det\left\{  
M^{(1)}\left[M^{(0)}\right]^{-1}\right\}\nonumber\\
&\times& 
\exp \frac{2e^2}{\pi}\intt \phi (x) \left[ \tilde E 
+\frac{1}{2}\Delta \phi (x)\right] 
\label{facinsta}
\end{eqnarray}
where the elements of the matrices 
 $N^{(\alpha)}$ and $M^{(\alpha)}$ are 
%\begin{eqnarray}
$N_{pp'}^{(\alpha)}=\intt \phi^{(\alpha)\dagger}_p 
\phi_{p'}^{(\alpha)}$ and 
%\nonumber\\
$M_{pp'}^{(\alpha)}=\intt 
\varphi^{(\alpha)\dagger}_p \varphi_{p'}^{(\alpha)}$.
%\label{mnmatr}
%\end{eqnarray}

Secondly, we are going to calculate the spectrum of $H(\tilde A;\mu)$ 
  in (\ref{hpmphinz}), with the boundary conditions discussed in the text. 
 Like in the $\mu=0$ case \cite{sawi92}, we shall try eigenfunctions 
bearing the form:
\begin{equation}
\phi^\pm_{n,m}=e^{(2n+1)\pi i/\beta}
e^{ i\bar h_1\mp i\mu)x_1}\xi_{n,m}^\pm (x_1)
\label{ansatzpsi}
\end{equation}

By plugging (\ref{ansatzpsi}) into the eigenvalue equation, we 
arrive at an 
harmonic oscillator eigenvalue problem, which can be solved 
 in the standard fashion. However, the functions $\phi^\pm$ in 
(\ref{ansatzpsi}) 
do not satisfy the right boundary conditions. 
 It is 
not difficult to see that:
%\begin{equation}
%\phi^\pm_{n,m}(x^0,x^1+L)=e^{-i\Phi x^0/\beta}e^{2\pi ieh_1\mp 
%i\mu L}\phi^\pm_{n+k',m}(x^0,x^1)
%\end{equation}

\begin{equation}
\hat\phi^\pm_{n,m} (x^0,x^1)=
\sum_{j=-\infty}^{\infty}e^{(2\pi ieh_1\mp i\mu L)j}
\phi^\pm_{n+jk',m}(x^0,x^1)
\label{eigentrue}
\end{equation}
 with $k'\equiv n_+-n_-=\Phi/2\pi$, 
 are the correct eigenfunctions, which do satisfy the right 
boundary conditions in (\ref{ferspbc}). We have used that   
the $\xi (x_1)$ functions in 
(\ref{ansatzpsi}) depend on $x^1$ and $n$ only through  
the combination $y=x^1+2\pi L(n+1/2-eh_0)/\Phi$.

The usual  harmonic oscillator 
quantisation condition for the $\xi^\pm_{n,m}$ states reads in this case:
\begin{equation}
\frac{\lambda^\pm L\beta}{\vert\Phi\vert}\pm\sgn\Phi=2m+1
\end{equation}
$m$ being an integer, with $m\geq 0$. Then, the eigenvalues for 
$\Phi\neq 0$ 
are independent on $\mu$, and they are: $\lambda=0$ with degeneracy 
 $k$, and $\lambda_m=2m\vert\Phi\vert/L\beta$ with degeneracy 2$k$. 
As we had anticipated, the zero modes appear with only one 
chirality, equal to the sign of $\Phi$. The eigenfunctions are 
those in (\ref{eigentrue}) and (\ref{ansatzpsi}) with
\begin{equation}
\xi^\pm_{n,m}=
H_m\left[\sqrt{\frac{\vert\Phi\vert}{L\beta}}y\right]\exp
\left[-\frac{\vert\Phi\vert}{2L\beta}y^2\right]
\end{equation}
where  $H_m$ are the Hermite polynomials. 
Once we know  the eigenvalues, we can calculate the determinant using 
 again  $\zeta$-function regularization and we get:
%\begin{equation}
%\zeta_H (s)=2k \left(\frac{L\beta}{2\vert\Phi\vert}\right)^s\zeta (s)
%\end{equation}
% $\zeta (s)$ being the Riemann zeta function. Upon recalling that 
%$\zeta (0)=1/2$ and $\zeta'(0)=-\log \sqrt{2\pi}$, we obtain:
\begin{equation}
{\det}' H(\tilde A;\mu)=\exp\left[-\left.\frac{d}{ds}\zeta_H (s)\right
\vert_{s=0}\right]=
\left(\frac{\pi L\beta}{\vert\Phi\vert}\right)^k
\label{detpatilde}
\end{equation}
 independent on $\mu$. 
If we concentrate only in the zero modes, that is, the $(n,m)$ 
eigenstates in (\ref{eigentrue}) 
with $m=0$, it turns out that they are already orthogonal and 
that their norms are independent on $\mu$:  
\begin{equation}
\vert\vert \phi_{n,0}\vert\vert^2=\left(\frac{\pi L\beta^3}
{\vert\Phi\vert}\right)^{1/2}
\label{normzeromod}
\end{equation}

As it was commented in the text, once we know the spectrum 
of the Dirac operator for 
 $\Phi\neq 0$ we could calculate the chiral condensates, 
which do not depend on the Green function 
 $G(x,y,eA;\mu)$, up to the phase factor. In particular, we have
\begin{equation}
 \langle \overline\psi (x) P_{\pm}\psi (x)\rangle_f=
\exp\left[i{\cal J}^{(1)}(A;\mu)\right]\sqrt{{\det}'H(A;\mu)} 
 \varphi^\dagger_1 (x)P_{\pm}\phi_1(x)
\end{equation}
where the superscript 1   indicates the sector 
with only one zero mode.

\end{document}